\documentclass[useAMS,usenatbib]{aa}
\usepackage{amsmath}
\usepackage{graphicx}
\usepackage{subfig}
\usepackage{float}
\usepackage{upquote}
\usepackage{footnote}
\usepackage[varg]{txfonts}
\def\jref@jnl#1{{\rm#1}}

\def\aj{\jref@jnl{AJ}}                   % Astronomical Journal
\def\araa{\jref@jnl{ARA\&A}}             % Annual Review of Astron and Astrophys
\def\apj{\jref@jnl{ApJ}}                 % Astrophysical Journal
\def\apjl{\jref@jnl{ApJ}}                % Astrophysical Journal, Letters
\def\apjs{\jref@jnl{ApJS}}               % Astrophysical Journal, Supplement
\def\ao{\jref@jnl{Appl.~Opt.}}           % Applied Optics
\def\apss{\jref@jnl{Ap\&SS}}             % Astrophysics and Space Science
\def\aap{\jref@jnl{A\&A}}                % Astronomy and Astrophysics
\def\aapr{\jref@jnl{A\&A~Rev.}}          % Astronomy and Astrophysics Reviews
\def\aaps{\jref@jnl{A\&AS}}              % Astronomy and Astrophysics, Supplement
\def\azh{\jref@jnl{AZh}}                 % Astronomicheskii Zhurnal
\def\baas{\jref@jnl{BAAS}}               % Bulletin of the AAS
\def\jrasc{\jref@jnl{JRASC}}             % Journal of the RAS of Canada
\def\memras{\jref@jnl{MmRAS}}            % Memoirs of the RAS
\def\mnras{\jref@jnl{MNRAS}}             % Monthly Notices of the RAS
\def\pra{\jref@jnl{Phys.~Rev.~A}}        % Physical Review A: General Physics
\def\prb{\jref@jnl{Phys.~Rev.~B}}        % Physical Review B: Solid State
\def\prc{\jref@jnl{Phys.~Rev.~C}}        % Physical Review C
\def\prd{\jref@jnl{Phys.~Rev.~D}}        % Physical Review D
\def\pre{\jref@jnl{Phys.~Rev.~E}}        % Physical Review E
\def\prl{\jref@jnl{Phys.~Rev.~Lett.}}    % Physical Review Letters
\def\pasp{\jref@jnl{PASP}}               % Publications of the ASP
\def\pasj{\jref@jnl{PASJ}}               % Publications of the ASJ
\def\qjras{\jref@jnl{QJRAS}}             % Quarterly Journal of the RAS
\def\skytel{\jref@jnl{S\&T}}             % Sky and Telescope
\def\solphys{\jref@jnl{Sol.~Phys.}}      % Solar Physics
\def\sovast{\jref@jnl{Soviet~Ast.}}      % Soviet Astronomy
\def\ssr{\jref@jnl{Space~Sci.~Rev.}}     % Space Science Reviews
\def\zap{\jref@jnl{ZAp}}                 % Zeitschrift fuer Astrophysik
\def\nat{\jref@jnl{Nature}}              % Nature
\def\iaucirc{\jref@jnl{IAU~Circ.}}       % IAU Cirulars
\def\aplett{\jref@jnl{Astrophys.~Lett.}} % Astrophysics Letters
\def\apspr{\jref@jnl{Astrophys.~Space~Phys.~Res.}}
\def\bain{\jref@jnl{Bull.~Astron.~Inst.~Netherlands}} 
\def\fcp{\jref@jnl{Fund.~Cosmic~Phys.}}  % Fundamental Cosmic Physics
\def\gca{\jref@jnl{Geochim.~Cosmochim.~Acta}}   % Geochimica Cosmochimica Acta
\def\grl{\jref@jnl{Geophys.~Res.~Lett.}} % Geophysics Research Letters
\def\jcp{\jref@jnl{J.~Chem.~Phys.}}      % Journal of Chemical Physics
\def\jgr{\jref@jnl{J.~Geophys.~Res.}}    % Journal of Geophysics Research
\def\jqsrt{\jref@jnl{J.~Quant.~Spec.~Radiat.~Transf.}}
\def\memsai{\jref@jnl{Mem.~Soc.~Astron.~Italiana}}
\def\nphysa{\jref@jnl{Nucl.~Phys.~A}}   % Nuclear Physics A
\def\physrep{\jref@jnl{Phys.~Rep.}}   % Physics Reports
\def\physscr{\jref@jnl{Phys.~Scr}}   % Physica Scripta
\def\planss{\jref@jnl{Planet.~Space~Sci.}}   % Planetary Space Science
\def\procspie{\jref@jnl{Proc.~SPIE}}   % Proceedings of the SPIE

\begin{document}

\title{Spiral arm pitch angle and galactic shear rate in N-body simulations of disc galaxies}
\author{Robert J.J. Grand \thanks{robert.grand.10@ucl.ac.uk} \and Daisuke Kawata\and Mark Cropper}
 
 \institute{Mullard Space Science Laboratory, University College London, Holmbury St. Mary, Dorking, Surrey, RH5 6NT}

%\pagerange{\pageref{firstpage}--\pageref{lastpage}} \pubyear{2012}

%\pagerange{\pageref{firstpage}--\pageref{lastpage}}\pubyear{2011}
%\def\LaTeX{L\kern-.36em\raise.3ex\hbox{a}\kern-.15em
 % T\kern-.1667em\lower.7ex\hbox{E}\kern-.125emX}
  
 \abstract{
Spiral galaxies are observed to exhibit a range of morphologies, in particular in the shape of spiral arms. A key diagnostic parameter is the pitch angle, which describes how tightly wound the spiral arms are. Observationally and analytically, a correlation between pitch angle and galactic shear rate has been detected. For the first time, we examine whether this effect is detected in N-body simulations by calculating and comparing pitch angles of both individual density waves and overall spiral structure in a suite of N-body simulations. We find that higher galactic shear rates produce more tightly wound spiral arms, both in individual mode patterns (density waves) and in the overall density enhancement. Although the mode pattern pitch angles by construction remain constant with time, the overall logarithmic spiral arm winds over time, which could help to explain the scatter in the relation between pitch angle versus shear seen from observations. The correlation between spiral arm pitch angle and galactic shear rate that we find in N-body simulations may also explain why late Hubble type of spiral galaxies tend to have more open arms.
}

  
\titlerunning{Pitch angle - shear rate relation in N-body simulations}

\label{firstpage}

\maketitle

\section{Introduction}

The morphology of spiral galaxies, as laid out in the Hubble classification \citep{H26}, can be broadly characterised by the tightness of spiral arm structure and the size of the central region or bulge. In this classification scheme, more tightly wound spiral arms are associated with large central mass concentrations. The strong correlation between central mass concentration and pitch angle predicted by modal density wave theory \citep[e.g.][]{LS64,RRS75,SG79,BL89} is in accordance with this. However, there are complications in the Hubble classification scheme insofar as that this relation was derived from optical information of galaxies only. The correlation is not observed in the near-infrared wavelengths \citep{dJ96,SJ98a,SJ98b}, and some observational studies in the infrared waveband highlight a difference in morphology from that seen in the optical \citep[e.g.][]{BB94,Th96,GP98}. Moreover, the correlation between Hubble type and pitch angle has been shown to be weak \citep{K81} and the model predictions from density wave theory for spiral arm properties have been shown to have systematic offsets to observations \citep{KH82}. 

Despite these uncertainties in the Hubble type-pitch angle relation, more recent observations have shown convincing evidence for a correlation between spiral arm pitch angle and the shear rate of differentially rotating discs of spiral galaxies. \citet{SBP05} derived shear rates from the rotation curves of a sample of several barred galaxies and used Fourier analysis to draw the spiral shape. They found evidence for the shear rate dependency of the spiral arm pitch angle. Because the rotation curve shape is determined by the mass distribution, this is essentially a correlation between the central mass concentration and spiral arm pitch angle. This survey was later extended and the conclusion strengthened by \citet{SBB06}. 

The shear rate-pitch angle correlation is also supported by the analytical work based on swing amplification theory \citep{GLB65,T81} by \citet{JT66} \citep[see also][]{F01}, which calculated the spatial distribution of the response of the density of the differentially rotating stellar disc to a large perturbing mass. They showed that the density enhancement in this context is predicted to show smaller pitch angles (hence a more tightly wound structure) with increasing amount of shear present.

While theoretical and observational studies provide evidence for the shear rate-pitch angle relation, it has yet to be explored in N-body simulations. In this paper, we aim to study this relation by running a suite of N-body simulations of varying shear rates. For the first time we investigate the pitch angles of individual spiral wave mode patterns in N-body simulations by isolating the spiral wave mode patterns from the system using the conventional spectrogram analysis \citep[e.g.][]{QDBM10,Se12,SoS12,MFQD12,RD11} and calculating the spiral phase of the $m$-th mode. We find that the discs of higher shear rate exhibit systematically smaller pitch angles than their lower shear rate counterparts, as predicted from the theoretical studies mentioned above. We also trace the overall spiral arm feature and measure its pitch angle as a function of time. The motivation for exploring this pitch angle behaviour is that we and other authors have found that the pattern speed of the spiral arms in N-body simulations and observed galaxies decreases with radius in a similar manner to the angular rotation velocity of the disc particles \citep{MRM05,MRM06,SW11,WBS11,GKC12,GKC11,NDH12,CQ12,BSW12}. Because the pattern speed decreases in this way, the pitch angle decreases with time and leads to transient and recurrent spiral arm features that are seen in many simulations \citep[e.g.][and references therein]{Se10,Se11}. The evolving nature of the pitch angle of winding spiral arm features can be compared to the observational work of \citet{SBB06} which measures the pitch angle and shear rate of many spiral galaxies and reveals several different observed pitch angles for a given shear rate. 

The paper is organised as follows. The simulations are described in Sec. 2, the analysis techniques laid out in Sec. 3 and the results are described in Sec. 4 and 5 in which we also explore some of the other parameter space apart from shear rate. The discussion is presented in Sec. 6, followed by the conclusions in Sec. 7.

\section{Simulations}

\begin{table*} 
\begin{tabular}{c c c c c c c c c c }
  \hline\hline
  Simulation & $M_d (\times 10^{10}\rm{M_{\odot}})$ & $R_d (\rm{kpc})$ & $M_{vir} (\times 10^{12}\rm{M_{\odot}})$ & $c$ & $\zeta$ & N $(\times 10^6)$ & $\epsilon$ (pc) & $M_{b} (\times 10^{10}\rm{M_{\odot}})$ & $b$ \\ 
   %\cline{2-2}
   \hline
   F & $5.0 $ & $3.5 $ & $1.5 $ & 15 & $0.40$ & 1 & 340 & $4.0 $ & $0.5$ \\
   Fa & $5.0 $ & $3.5 $ & $1.5 $ & 15 & $0.40$ & 5 & 340 & $4.0 $ & $0.5$ \\
   Fb & $5.0 $ & $3.5 $ & $1.5 $ & 15 & $0.40$ & 5 & 200 & $4.0 $ & $0.5$ \\
   Fc & $5.0 $ & $3.5 $ & $1.5 $ & 15 & $0.40$ & 5 & 90 & $4.0 $ & $0.5$ \\ 
   F2 & $2.5 $ & $3.5 $ & $1.5 $ & 15 & $0.20$ & 1 & 270 & $4.0 $ & $0.5$ \\     
   F3 & $5.0 $ & $3.5 $ & $0.75 $ & 15 & $0.58$ & 1 & 340 & $2.5 $ & $0.5$ \\
   K & $5.0 $ & $3.5$ & $0.1 $ & 15 & $0.27$ & 1 & 340 & $10.0 $ & $0.01$ \\
   R & $1.0 $ & $3.5$ & $2.5 $ & 5 & $0.40$ & 1 & 200 & - & - \\
   R2 & $5.0 $ & $3.5$ & $1.5 $ & 20 & $0.53$ & 1 & 340 & - & - \\
   R3 & $5.0 $ & $3.5$ & $2.0 $ & 20 & $0.46$ & 1 & 340 & - & - \\
   R4 & $5.0 $ & $3.5$ & $3.0 $ & 10 & $0.83$ & 1 & 340 & - & - \\
   \hline
\end{tabular} 
\caption{Table of simulation parameters. Column (1) simulation name (2) disc mass (3) scale length (4) virial mass (5) NFW concentration parameter (6) disc to halo mass ratio within two scale lengths (7) number of particles (8) softening length (9) bulge mass (10) bulge compacting factor.}
\label{Inip}
\end{table*} 

The simulations in this paper are performed with a hierarchical Tree N-body code GCD+ \citep{KG03,KOG13}. We run a suite of simulations, each of which consists of a spherical static dark matter halo (and a spherical static stellar bulge component in some cases) and a live stellar disc. The halo and bulge are static rather than live in order to facilitate greater control of the experimental scenarios. A live halo/bulge component will complicate the evolution of the stellar disc with effects such as scattering and heating, and may even act as large perturbing masses that greatly disturb the disc if the mass resolution for the dark matter is too small \citep{DO12}. These are unwanted effects, and because the focus of this study is on the stellar disc component only, we have elected to model the external components with static potentials. 

The dark matter halo density profile follows that of \citet{NFW97} with the addition of an exponential truncation term \citep{RA11}:

\begin{eqnarray}
\rho _{\rm dm} = \frac{3 H_0^2}{8 \pi G} \frac{\Omega _0-\Omega_b}{\Omega_0 } \frac{\rho _{\rm c}}{ cx(1+cx)^2}\exp(-x^2) ,
\label{eq1}
\end{eqnarray} 
where $\rho _{\rm c}$ is the characteristic density described by \citet{NFW97}, the concentration parameter, $c = r_{\rm 200}/r_{\rm s}$, and  $x= r/r_{\rm 200}$. The truncation term, $\exp(-x^2)$, is introduced in our initial condition generator for a live halo simulation. Although we use a static dark matter halo in this paper, we retain the profile of equation (\ref{eq1}) because this term does not change the dark matter density profile in the inner region, which is the focus of this paper. The scale length is $r_{\rm s}$, and $r_{\rm 200}$ is the radius inside which the mean density of the dark matter sphere is equal to 200$\rho _{\rm crit}$ (where $\rho _{crit} = 3 H_0^2 / 8 \pi G$; the critical density for closure):

\begin{eqnarray}
r _{200} = 1.63 \times 10^{-2} \left( \frac{M_{vir}}{h^{-1} M_{\odot}} \right)^{\frac{1}{3}} h^{-1} \rm kpc,
\end{eqnarray} 
where $M_{vir}$ is the virial mass of the galaxy.

We assume $\Omega _0 = 0.266$, $\Omega_b = 0.0044$ and $H_0=71$ $\rm km$ $\rm s^{-1}$ $\rm Mpc^{-1}$. 

% Rotation curve (Initial conditions)
\begin{figure}
\centering
\includegraphics[scale=0.43]{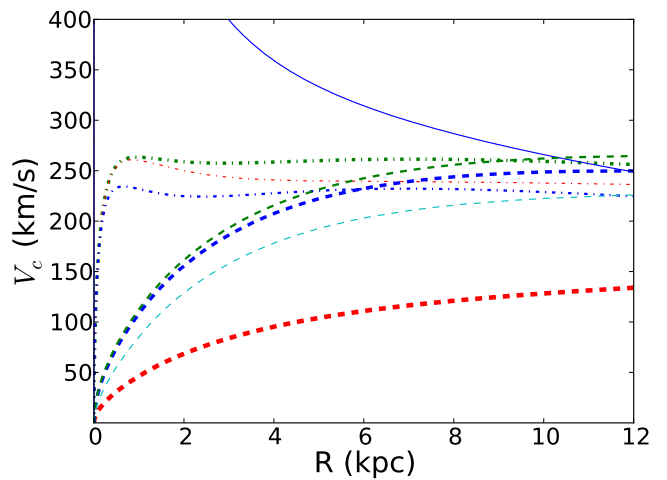}
\caption[]
{The circular velocity at $t = 0$ for simulation R (very thick dashed red), R2 (thick dashed blue), R3 (medium dashed green), R4 (thin dashed cyan), F (thick dot-dashed green), F2 (thin dot-dashed red), F3 (medium dot-dashed blue) and K (solid blue).}
\label{3v}
\end{figure}

\begin{figure}
\centering
\includegraphics[scale=0.43]{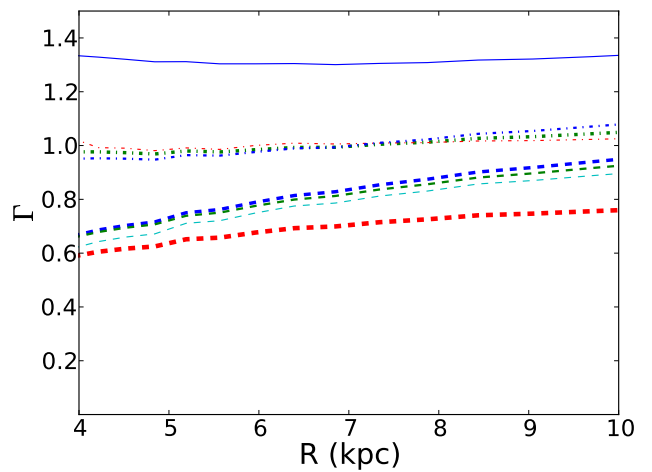}
\caption[]
{Galactic shear rate, $\Gamma$, for all simulations. Colours are the same as Fig. \ref{3v}. Note the reduced radial range compared to Fig. \ref{3v}}
\label{shear}
\end{figure}

\begin{figure}
\centering
\includegraphics[scale=0.43]{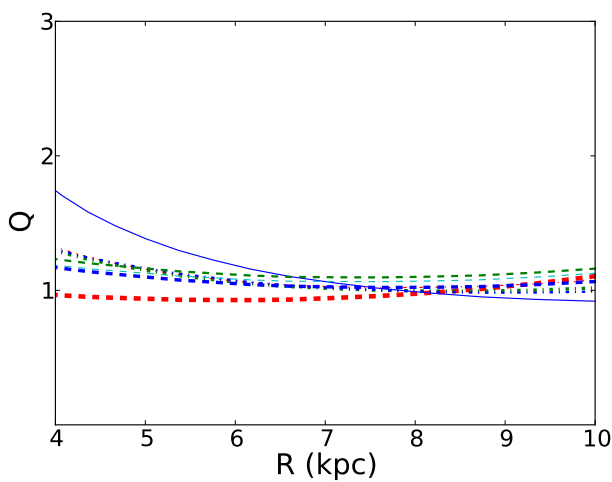}
\caption[]
{Toomre stability parameter, $Q$, at $t = 0$ for all simulations. Colours are the same as Fig. \ref{3v}.}
\label{Qprof}
\end{figure}

The spherical static stellar bulge component is modelled by the Hernquist profile \citep{H90}, which is described by:

\begin{eqnarray}
\rho _b (r) = \frac{M_b}{2\pi}\frac{a}{r}\frac{1}{(r+a)^3} ,
\label{hernb}
\end{eqnarray} 
where $M_b$ is the total bulge mass and $a$ is the scale length. The scale length is set to the effective radius, $R_e = 1.8153 a$. We apply a compacting factor, $b$, to scale from the empirical relation of the bulge effective radius \citep{SMW03}: 

\begin{eqnarray}
R_{e} = 4.16 \left( \frac{M_b}{10^{11} \rm{M_{\odot}}}\right).
\label{Reff}
\end{eqnarray} 
Hence, the resultant scale-length is defined by $a = b R_{e} / 1.8153$.

The stellar disc is assumed to follow an exponential surface density profile:

\begin{eqnarray}
\rho _{\rm d,*} = \frac{M_{\rm d,*}}{4 \pi z_{\rm d,*} R_{\rm d,*}^2} {\rm sech}^2 \left(\frac{z}{z_{\rm d,*}}\right) {\rm exp}\left(-\frac{R}{R_{\rm d,*}}\right).
\end{eqnarray} 
The fiducial number of disc particles used is $N=1 \times 10^6$. Numbers of this order are reported to be sufficient to minimise numerical heating \citep{Fu11}. Although larger particle numbers reduce numerical heating further, we note that the effect is always present (i.e. it does not disappear at a particular resolution), and that a compromise between parameter space and resolution must be made for suites of simulations such as the one presented in this study. 

We apply a fixed softening length, $\epsilon$, for star particles with the spline softening suggested by \citet{PM07}. The softening length\footnote{It should be noted that we define the softening length at which the softening kernel function is truncated. Therefore, our softening length value is typically a factor $\sim 3$ larger than the traditional definition: to translate our softening lengths to the traditional values, our value should be divided by 3.} is dependent on the particle mass, therefore the base value of $\epsilon = 340$ pc for the particle mass, $m_p = 5 \times 10^4 \rm{M_{\odot}}$ varies between simulations that have different particle masses. The model parameters for the simulations are summarised in Table \ref{Inip}, and the rotation curves are shown in Fig. \ref{3v}. 

There are three groups of rotation curves. Simulation group R (R, R2, R3, R4) has a rising rotation curve. Simulation R is an extreme case, where we set a large halo mass with a low concentration parameter, $c$, in order to extend mass to the outer regions of the disc. Because of such a low concentration of dark matter mass in the central region, the disc mass must be lowered in order to prevent a bar from forming \citep{OP73}. In this way, we avoid the added complication of the bar component and restrict the study to spiral galaxies only. Simulations R2, R3 and R4 are less extreme cases, which explore intermediate shear rates and different disc to halo mass ratios. To produce the flat (simulations F, Fa, Fb, Fc, F2 and F3) and Keplerian-like (simulation K) rotation curves, a bulge component is included. For simulation K, this is a very compact and massive bulge. Although this case is unrealistic, we include it in order to emphasise the effect of galactic shear on spiral morphology. 

The radial profile of the galactic shear rate at $t = 0$, given by:

\begin{eqnarray}
\Gamma = 1 - (R/V_c)(dV_c/dR), 
\label{sheqn}
\end{eqnarray}
for each simulation is shown in Fig. \ref{shear}. This suite of simulations represents a range of shear rates, which is the principal variable we want to investigate. However, there are other parameters that may affect the pitch angle, such as the disc-halo mass ratio, $\zeta$, softening length, $\epsilon$, and resolution. We also explore these parameters, mainly with simulation group F.

We set the initial Toomre stability parameter, $Q$, for all our simulations to approximately $1$ over the radial range $4 < R < 10$ kpc, which allows the spiral structure to grow\footnote{Each simulation shows a rise in the radial $Q$ profile over time, owing to the heating by spiral arm structure \citep{Fu11}.}. The radial dependence in shown in Fig. \ref{Qprof}.

\section{Method of Analysis}

Here we present the analysis method of our two techniques for measuring pitch angles: mode pattern analysis and direct spiral arm peak trace method. An important difference between these techniques is that the mode pattern analysis assumes that the spiral arms are constructed by one or multiple density waves of mode, $m$, which describe patterns of $m$ spiral arms with a constant pitch angle. The direct spiral arm peak trace method does not assume any theory, but simply analyses the pitch angle of the overall spiral arm feature. The distinction between these two methods is that while both characterise the spiral arm as a logarithmic spiral of fixed pitch angle at all radii of interest, in the direct method the pitch angle and amplitude of the spiral arm changes with time. However, in the mode analysis, changes in the spiral arm (in particular the winding) may only occur through the changing \emph{superposition} of the various mode patterns present. 

Before we describe these two analysis techniques, we define the pitch angle which we will use with both. Given the positional information ($R,\theta$) of a density enhancement, we can fit logarithmic spiral arms, described by:

\begin{eqnarray}
\theta = B \ln R + C,
\label{logs}
\end{eqnarray} 
where $\theta$ is the azimuth coordinate, $R$ is the radial coordinate and $B$ and $C$ are constants. Logarithmic spirals have pitch angles, $\phi$, given by \citep{BT08}:

\begin{eqnarray}
 \tan{\phi} = \frac{\Delta R}{d_{\theta}},
\label{pa}
\end{eqnarray} 
where the distance, $d_{\theta}$, is the spatial distance of the density enhancement in the azimuthal direction defined as $d_{\theta} = R\Delta \theta$. The pitch angle of a logarithmic spiral is constant with radius. The next step is to recover the positional information ($R,\theta$) required to apply the logarithmic chi-squared fitting using equation (\ref{logs}) and calculate the pitch angle of the fit using equation (\ref{pa}).

\subsection{Mode pattern analysis method}

By construction, a wave mode pattern has a constant pattern speed, $\Omega ^m _p$. Therefore the shape of a wave mode pattern is time independent \emph{i.e.} the pitch angle is constant over time. In this analysis, we focus on strong patterns because their behaviour is most evident. In order to find patterns of significant amplitude, we first search for dominant modes \emph{i.e.} wave modes of $m$ spiral arms that exhibit large amplitudes. The amplitude of a given wave mode, $m$, is calculated from the quantities:

\begin{eqnarray}
W_c^m(R,t) = \sum_i^N{\cos (m\theta_i)}, \nonumber \\
W_s^m(R,t) = \sum_i^N{\sin (m\theta_i)},
\label{W(t)}
\end{eqnarray} 
where $\theta _i$ is the azimuthal angle between the radial vector of the particle and a common reference vector. The amplitude is then calculated as:

\begin{eqnarray}
A^m(R,t) = (W_c^m(R,t)^2 + W_s^m(R,t)^2)^{1/2}.
\label{Ampli_t}
\end{eqnarray} 
The mean amplitude in a radial range $4-10$ kpc is calculated using equation (\ref{Ampli_t}) for modes $m = 1 - 7$ over the entire 2 Gyr of the evolution for each simulation. This is shown in the top row of Fig. \ref{pspec}. In each simulation, prominent modes are identified for analysis. We aim to extract the positional information of the patterns. The adopted procedure is to compute their power spectra by taking the Fourier transform of the time sequence of each component in equation (\ref{W(t)}) \citep{QDBM10}:

\begin{eqnarray}
\tilde{W}_c^m(R,\omega) = \int_{T_1}^{T_2}{W_c^m(R,t)e^{i\omega t} h(t)} dt, \nonumber \\
\tilde{W}_s^m(R,\omega) = \int_{T_1}^{T_2}{i W_s^m(R,t)e^{i\omega t} h(t)} dt,
\label{W(w)}
\end{eqnarray} 
where $h(t)$ denotes the Hanning function used to reduce the aliasing. $T_1$ and $T_2$ denote the beginning and end of the time window of the Fourier transform. This is chosen to be at around a relatively late epoch of the simulation (when the system is more stable) and is centred around a peak of the most dominant mode present in each case. It spans $\Delta t = 256$ Myr, which is a typical life time of a spiral arm as shown in the next section.

The amplitude in each frequency as a function of radius is then calculated via:

\begin{eqnarray}
A^m(R,\omega) = (\tilde{W}_c^m(R,\omega)^2 + \tilde{W}_s^m(R,\omega)^2)^{1/2}.
\label{Ampli}
\end{eqnarray} 
Because simulations generally possess several patterns for a given mode that can overlap in radius \citep[e.g. see Fig. 4 of][]{RD11}, care must be taken when computing the spiral phase of a pattern. In this technique each wave mode pattern is characterised by a pattern speed given by $\Omega ^m _p = \omega / m$, which is \emph{constant} over radius. Individual patterns should be selected by isolating a horizontal ridge (a single pattern speed) over a radial range where the signal significantly stands out from the noise. In each of the galaxies, we focus on the most dominant patterns and look at the three quantities, $\tilde{W}_c^m(R,\omega)$, $\tilde{W}_s^m(R,\omega)$ and $A^m(R,\omega)$ on the real and imaginary axis for each radial pixel in a ridge. We then calculate the real spiral arm phase position within the domain $0$ to $2 \pi$ as:

\begin{figure*}[!htbp]
\begin{center}
 % \hspace{-10.00mm} 
   \subfloat{\includegraphics[scale=0.29]{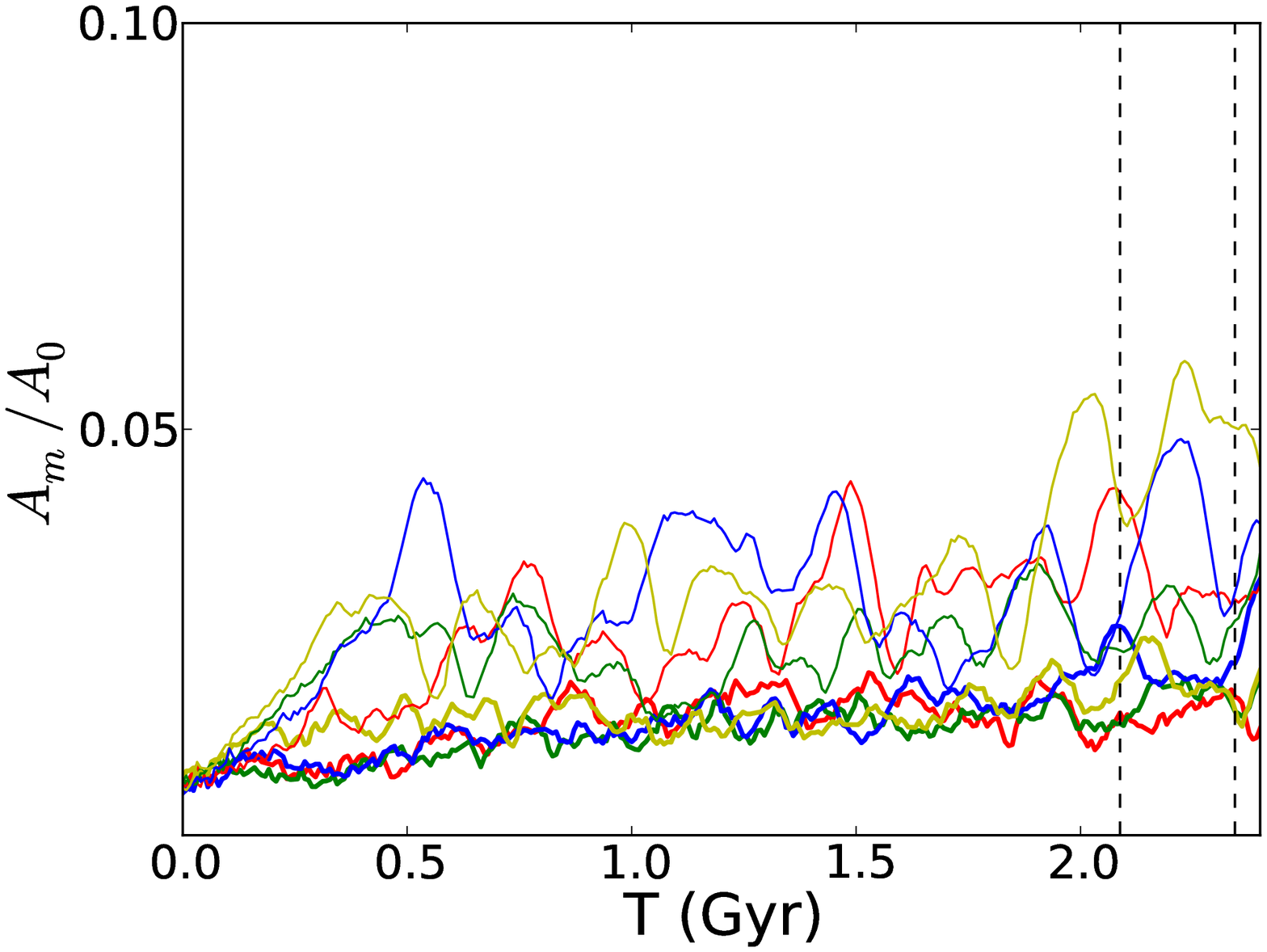}} 
  \subfloat{\includegraphics[scale=0.29] {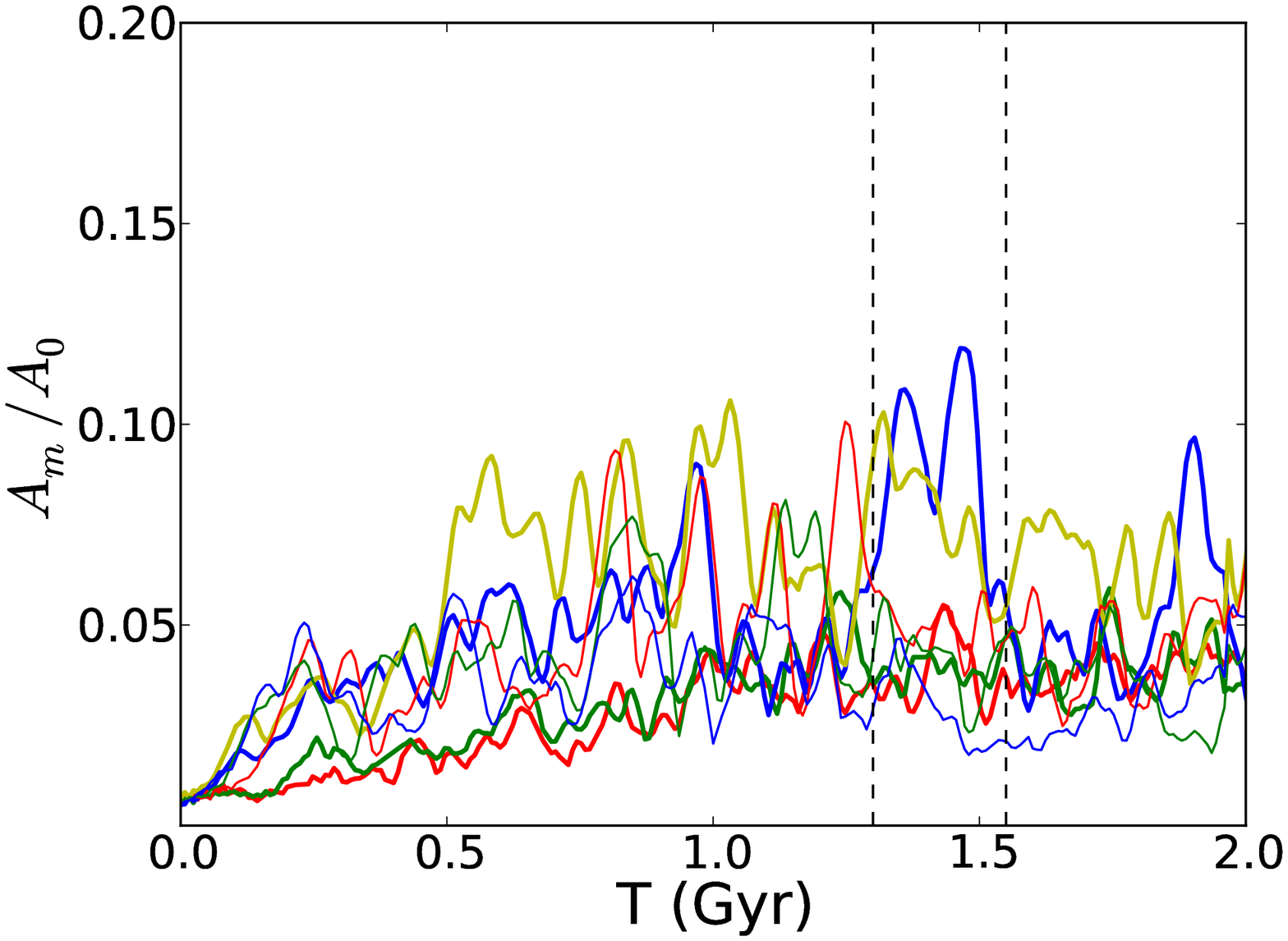}} 
  \subfloat{\includegraphics[scale=0.29] {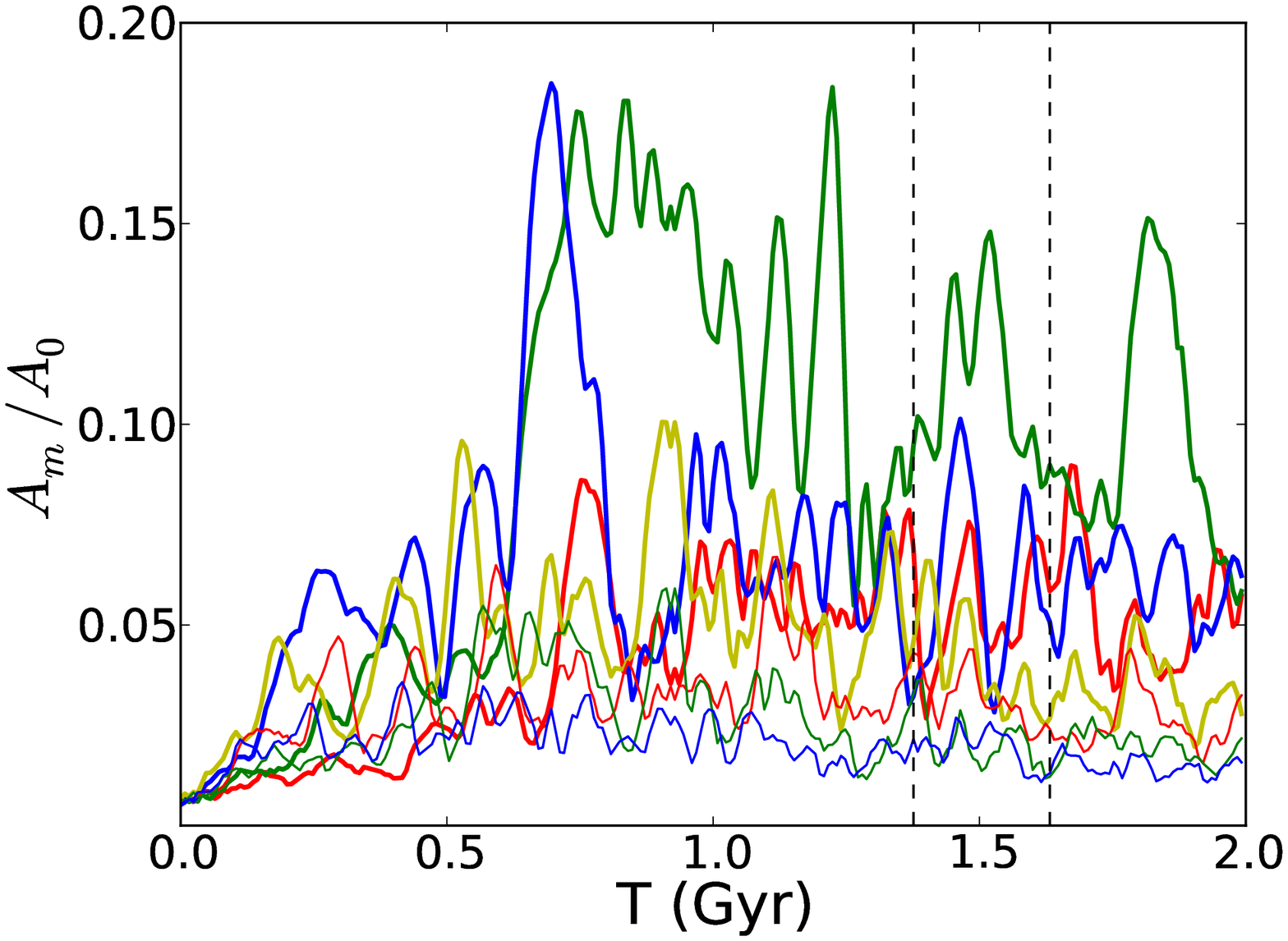}} \\
  \subfloat{\includegraphics[scale=0.21]{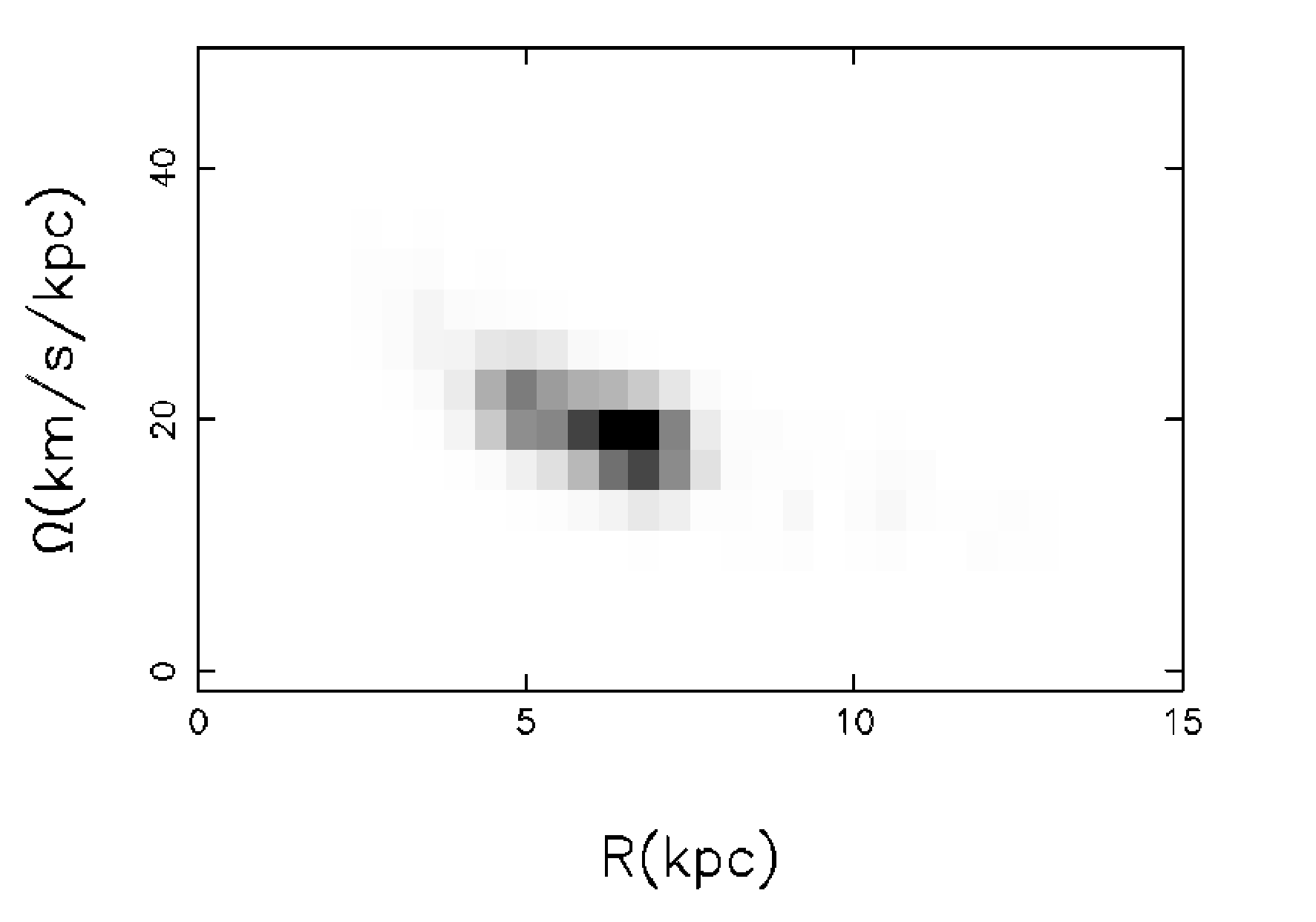}} 
  \subfloat{\includegraphics[scale=0.21] {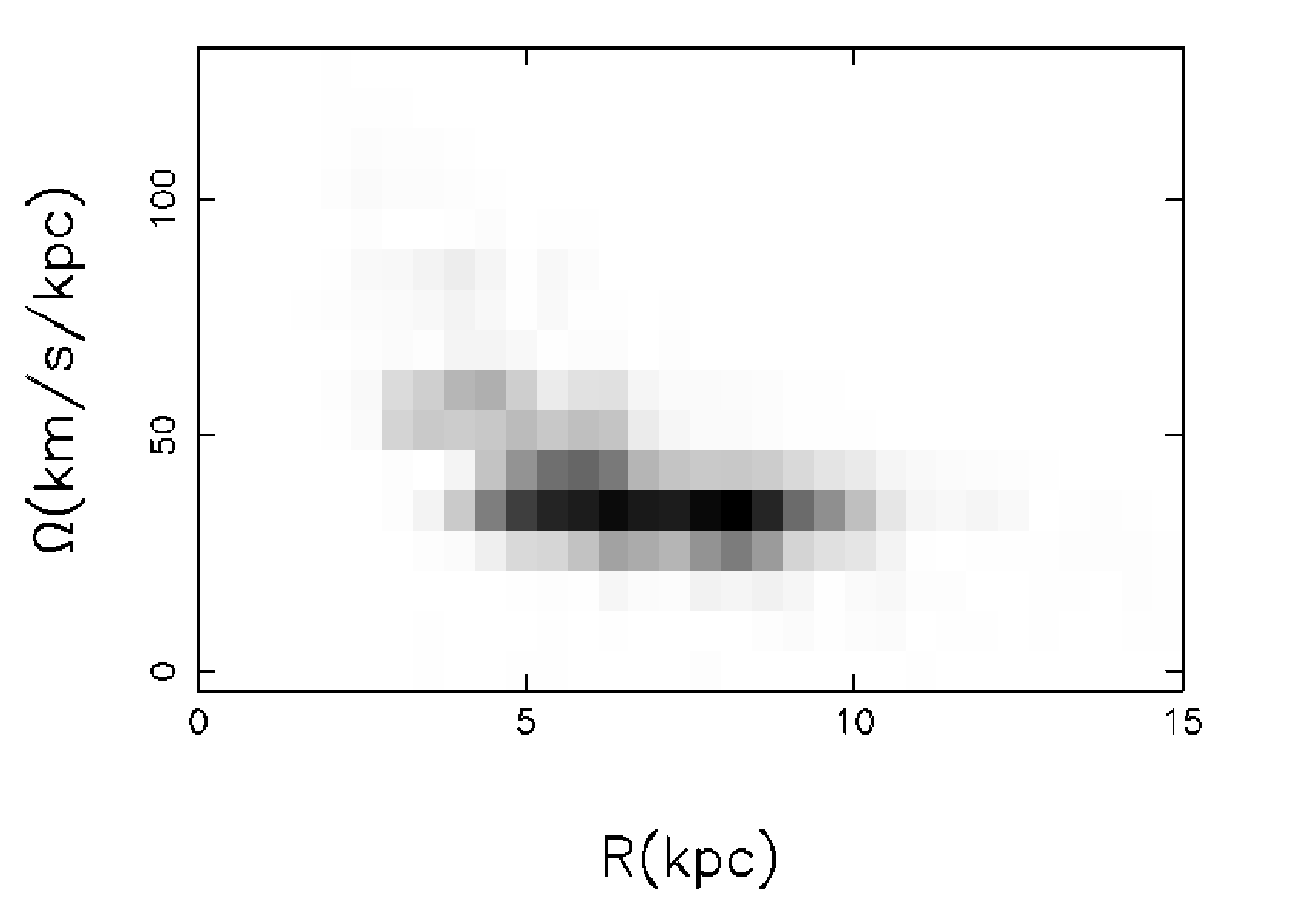}}
  \subfloat{\includegraphics[scale=0.21] {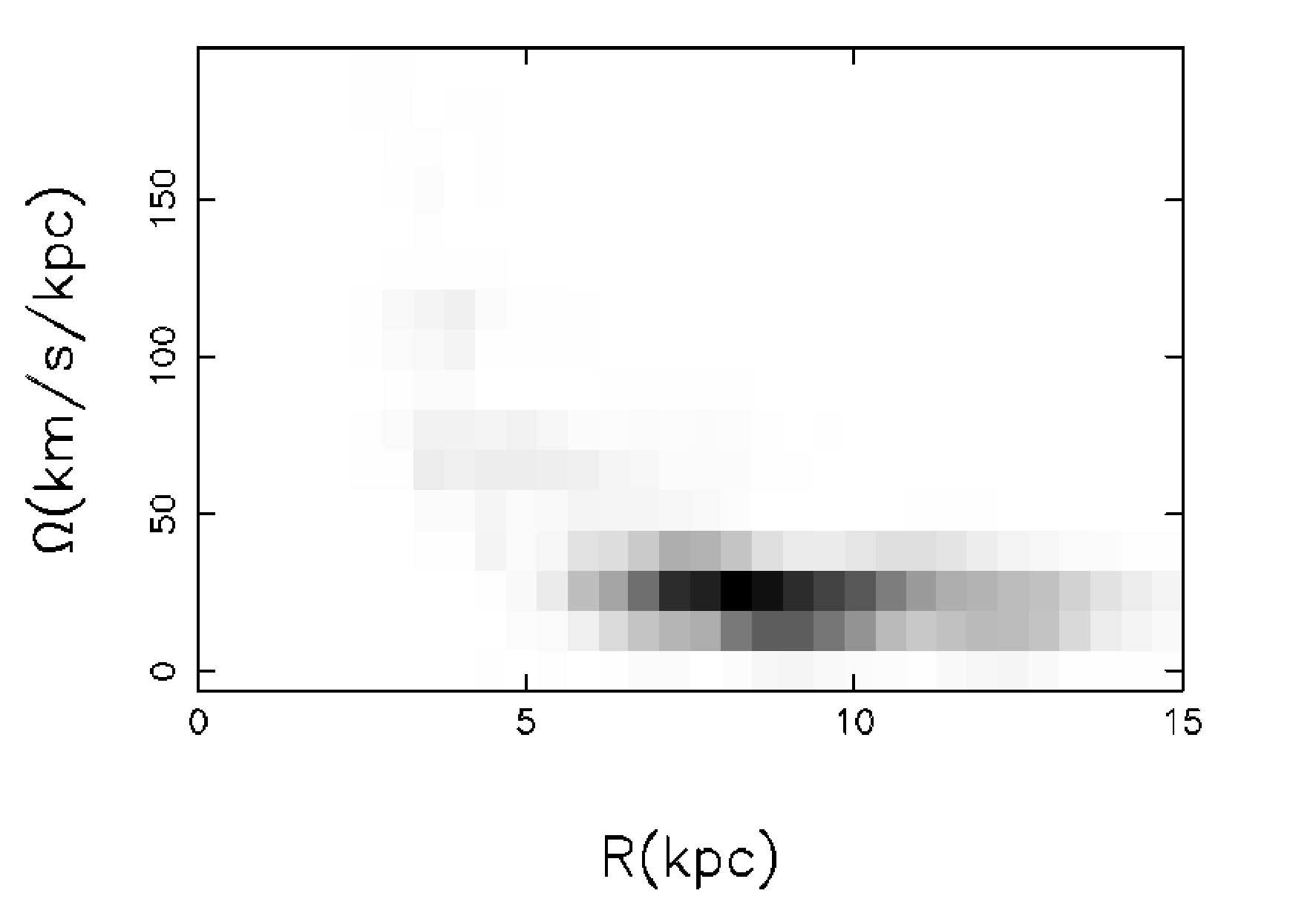}} \\
%  \hspace{-4.00mm}
  \subfloat{\includegraphics[scale=0.31]{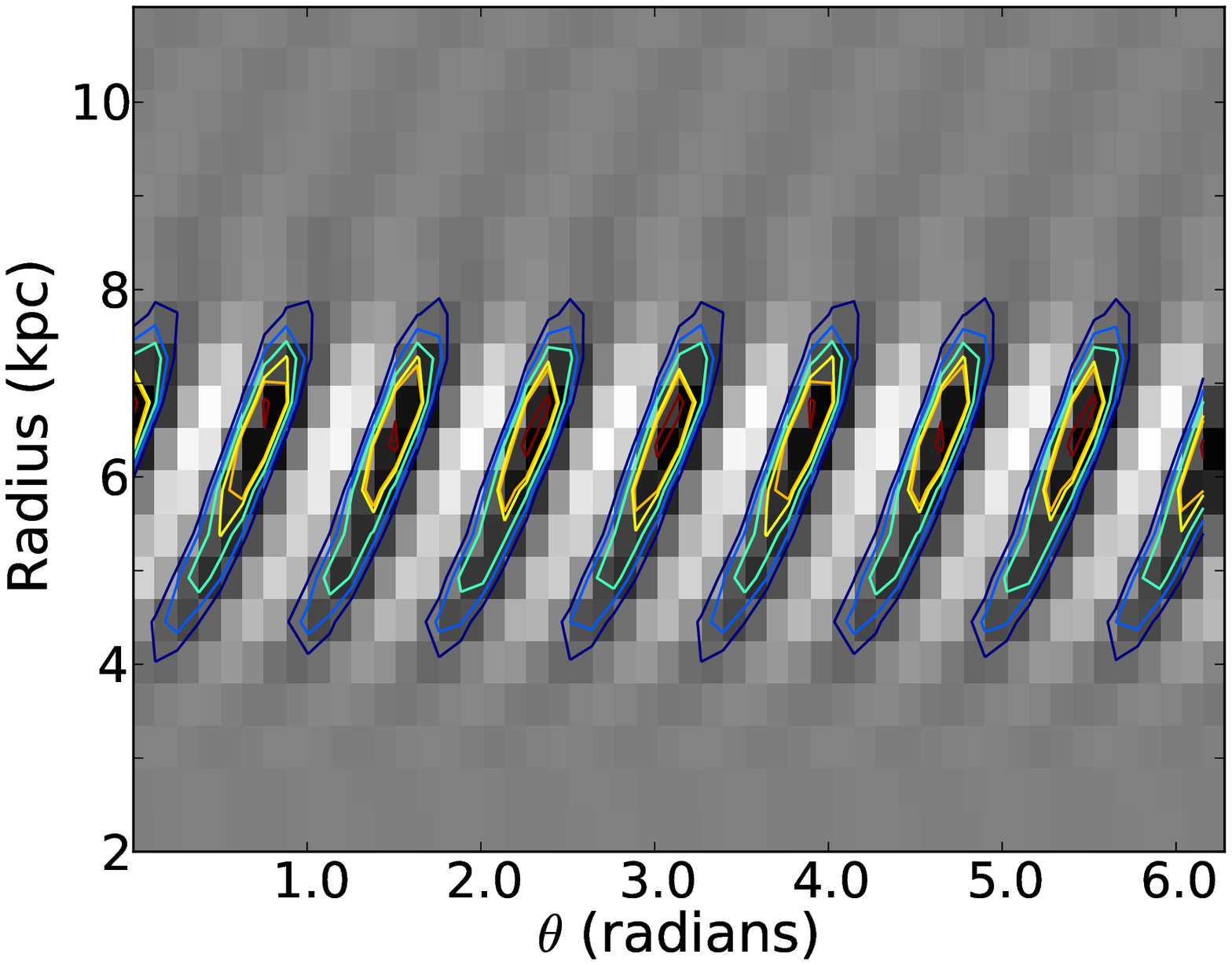}} 
  \subfloat{\includegraphics[scale=0.31] {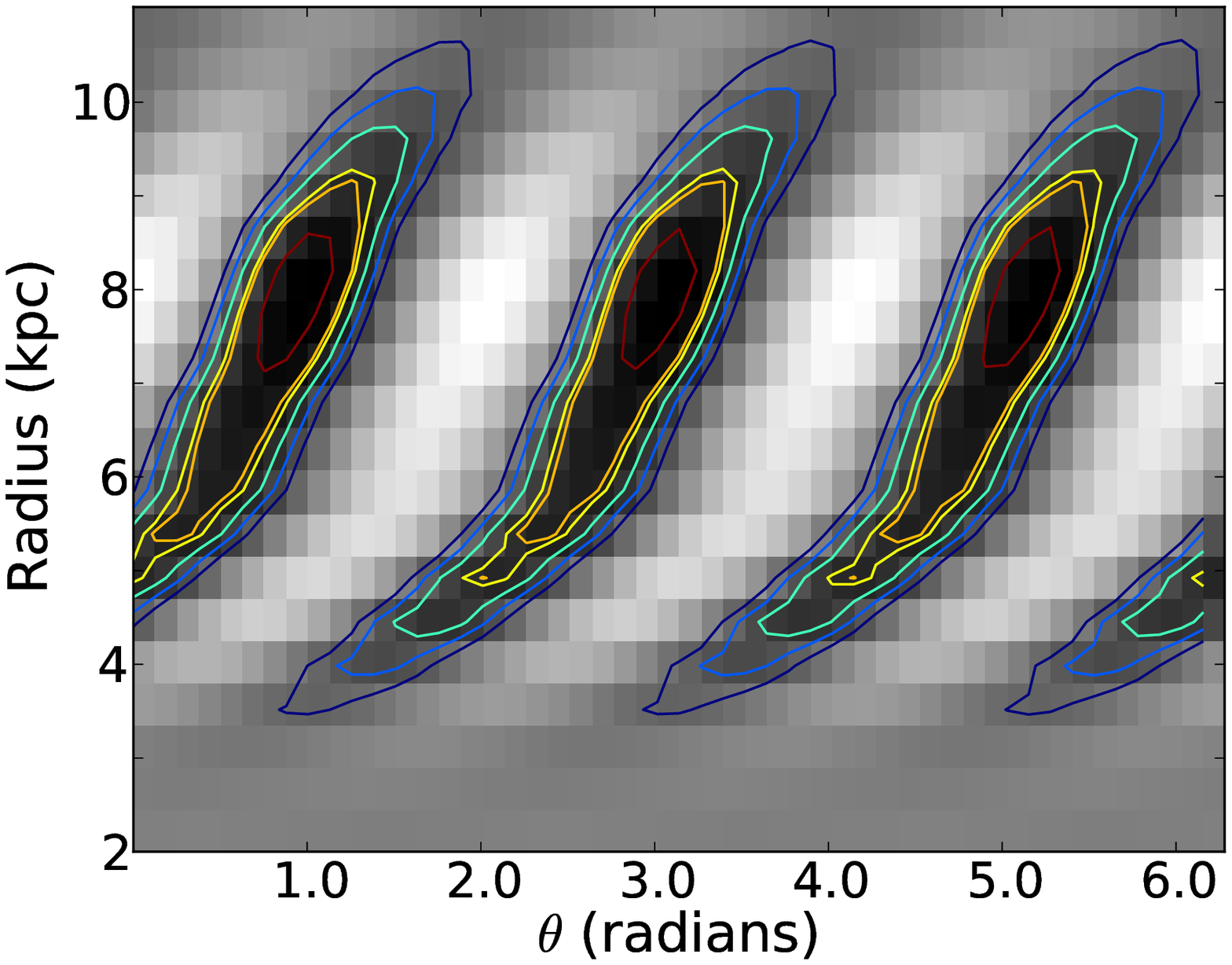}}
  \subfloat{\includegraphics[scale=0.31] {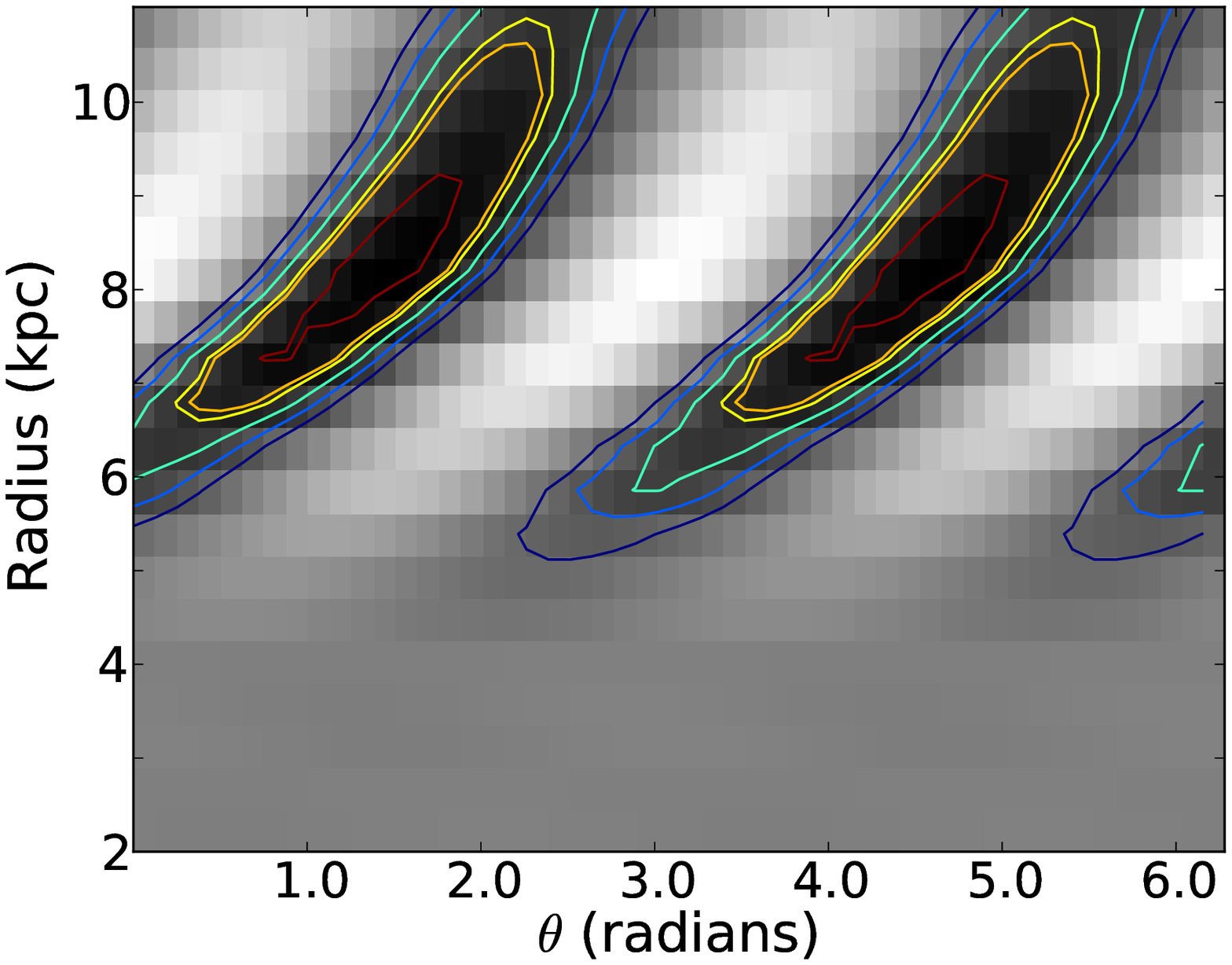}} \\  
  
 \subfloat{\includegraphics[scale=0.29]{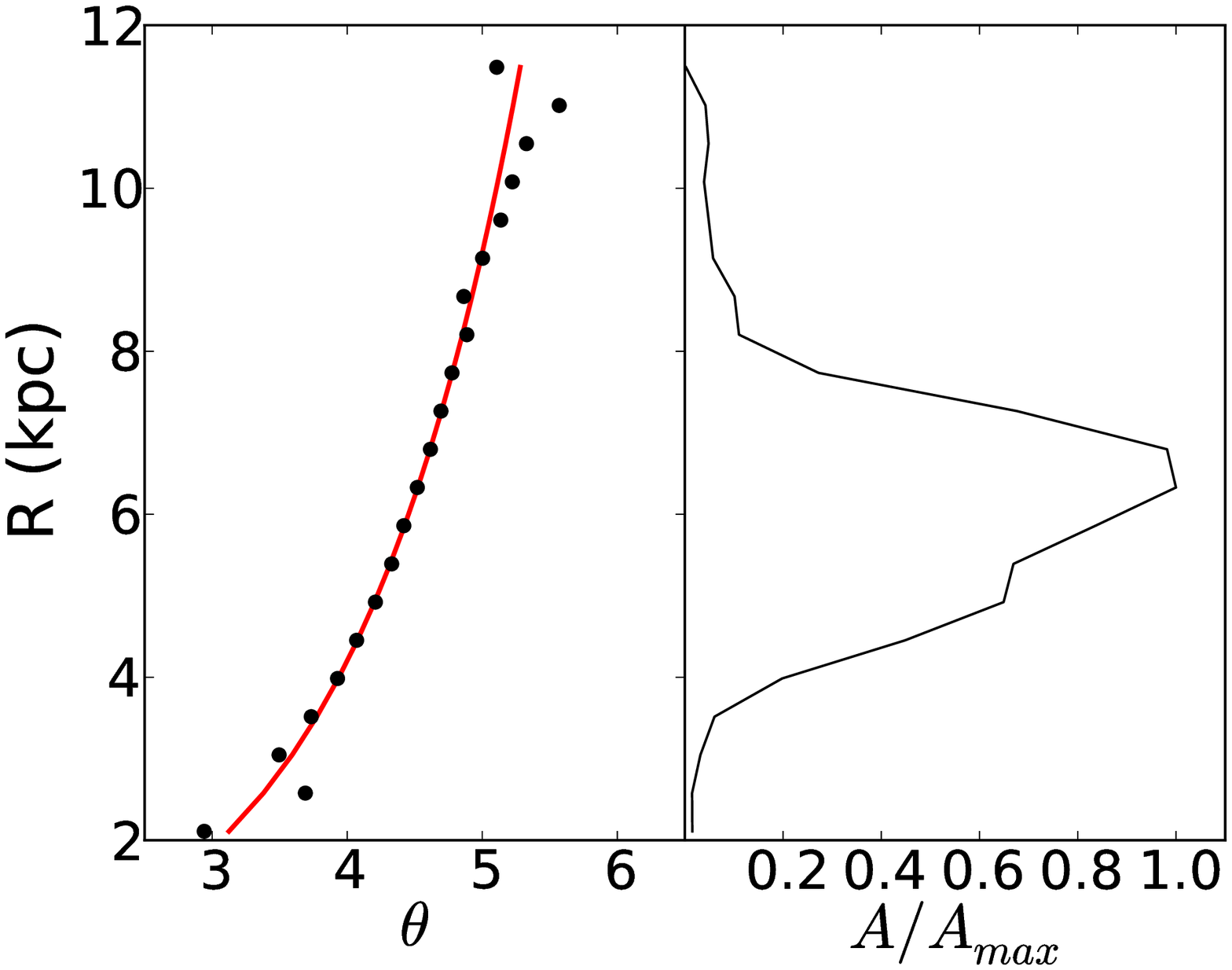}}
   \subfloat{\includegraphics[scale=0.29] {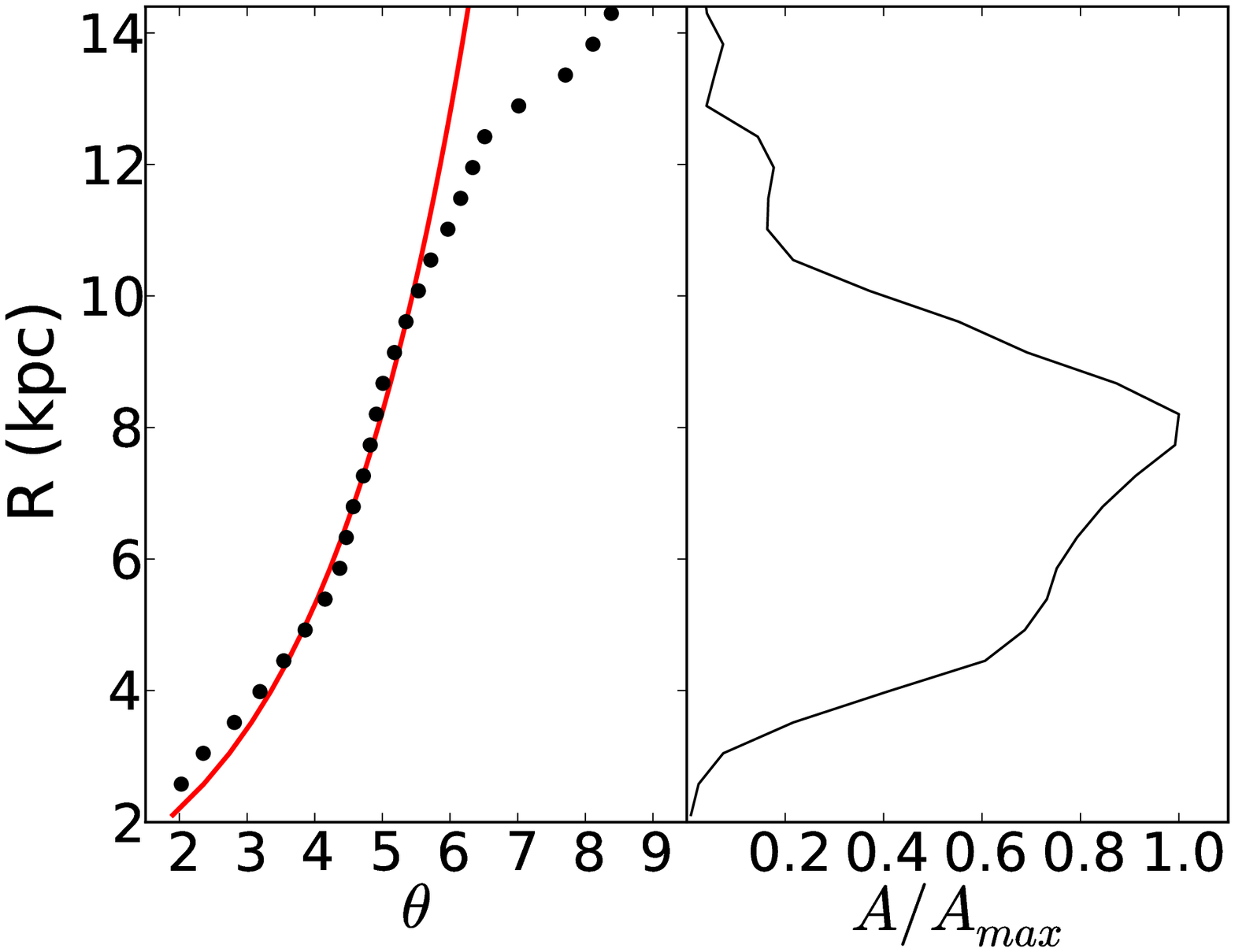}} 
  \subfloat{\includegraphics[scale=0.29] {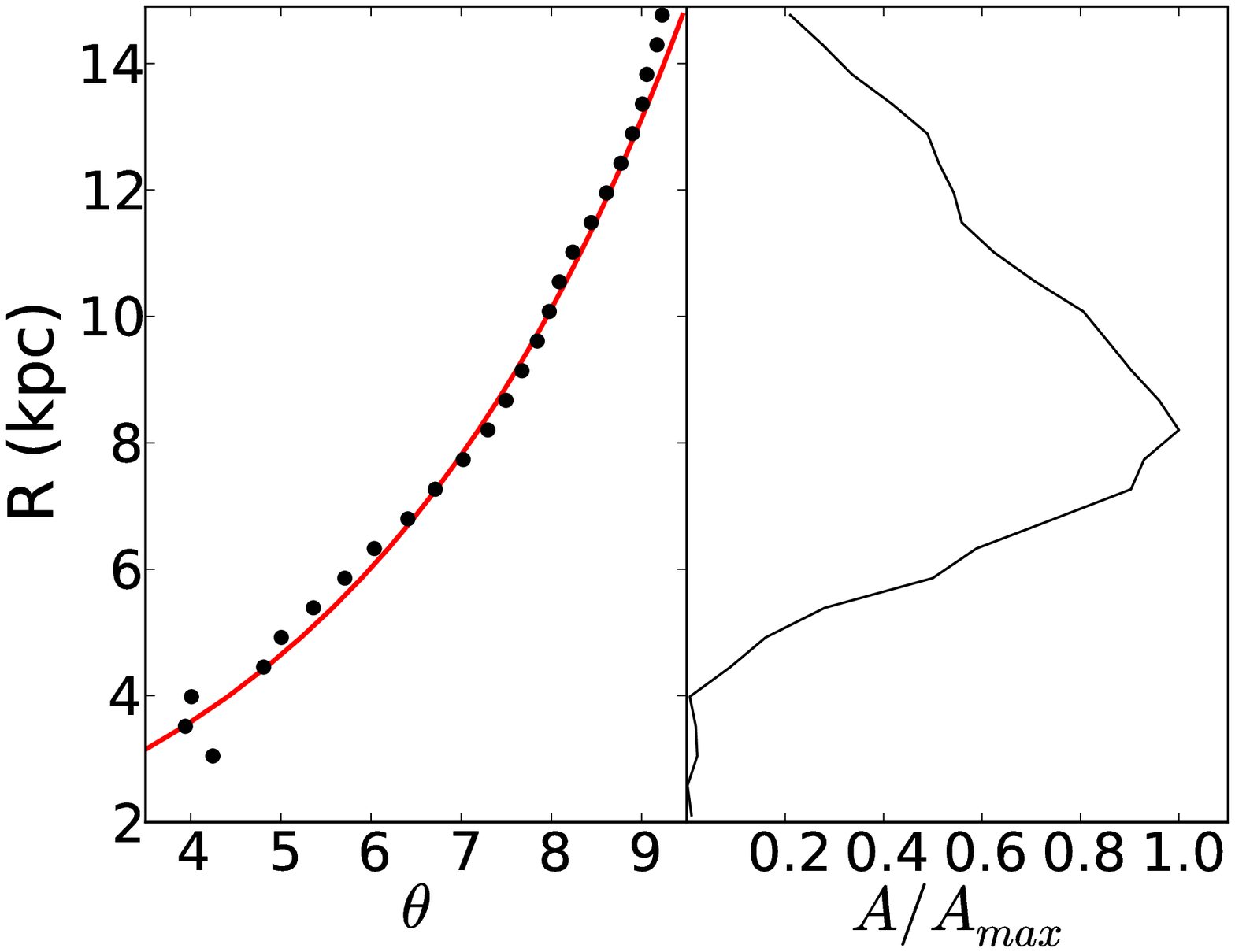}} \\

\caption[]{\emph{Top row}: The amplitudes calculated from equation (\ref{Ampli_t}) and averaged over a radial range of 4 - 10 kpc of spiral modes $m = 1$ (thick red), 2 (thick green), 3(thick blue), 4(thick yellow), 5 (thin red) 6 (thin green), 7 (thin blue) and 8 (thin yellow) normalised to the axisymmetric $m=0$ mode, as a function of time for simulations R (left), F (middle) and K (right). Vertical dashed lines represent the time window of a Fourier transform applied in Section 3.1. \emph{Second row} Power spectra calculated from equation (\ref{Ampli}) of simulation R for the $m=8$ mode (left), F for the $m=3$ mode (middle) and K for the $m=2$ mode (right). Prominent ridges (dark pixels) span between 4 - 10 kpc in most cases. \emph{Third row}: In polar coordinates, the density map of the dominant density wave mode pattern selected from rows of $\Omega ^m _p = 18$, $30$ and $24$ $\rm{km}$ $\rm{s^{-1}}$ $\rm{kpc^{-1}}$ for simulations R, F and K respectively. White regions indicate areas of low density and black regions indicate areas of high density. Contours emphasis the highest density regions. \emph{Bottom panels}: Dominant mode pattern positions (black points) calculated from equation (\ref{Sphase}) in the azimuth-radius plane for the corresponding patterns in the row above. The red lines show the lines of best fit for each pattern. The right side of each panel shows the radial amplitude profile, which is used to weight the fitting.}
\label{pspec}
\end{center}
\end{figure*} 

\begin{eqnarray}
\theta _p (R,\omega) = \frac{\theta^m_{sp} (R,\omega)}{m} = \frac{1}{m} \arctan \left(\frac{\tilde{W}_s^m(R,\omega)}{\tilde{W}_c^m(R,\omega)}\right),
\label{Sphase}
\end{eqnarray} 
where $\theta^m_{sp} (R,\omega)$ is the spiral phase of the pattern at each radial bin, which is retrieved by considering only the Fourier coefficients of a single $\omega$. Because this quantity spans a domain of $2 \pi m$, the spiral phase, $\theta^m_{sp} (R,\omega)$, is divided by $m$ in order to yield the real phase position of the wave mode pattern as a function of radius. This provides the azimuthal and radial values required for the calculation of the pitch angle using equations (\ref{logs}) and (\ref{pa}).

\subsection{Direct spiral arm peak trace method}

The method we use to trace the spiral arm peak position directly is a particle density weighting method, in which we select a point near the spiral arm of interest at some start radius ($\sim 5$ kpc), define an azimuth range that encapsulates the width of the spiral arm and weight by particle density to find the peak position \citep[see][for more details]{GKC12}. This is iterated over a radial range until the spiral arm peak position is drawn out. Several spiral arms are traced over a range of snapshots between $1$ and $2$ Gyr of the simulation evolution. Spiral arms are only traced when they show a single density peak over azimuth for each radius in the radial range chosen for fitting. The pitch angles are then calculated using equations (\ref{logs}) and (\ref{pa}).

We remind the reader that this pitch angle is derived from the spiral arm line that traces out the overall density enhancement directly, which varies with time. This is different from the \emph{time independent} pitch angle calculated from the positional information of the wave mode patterns derived from the power spectra (see Section 3.1). The latter bears the assumption of a density wave of constant pattern speed and fixed pitch angle, whereas the former bears no assumptions at all.

\begin{figure}[!h]
\begin{center}
\includegraphics[scale=0.42]{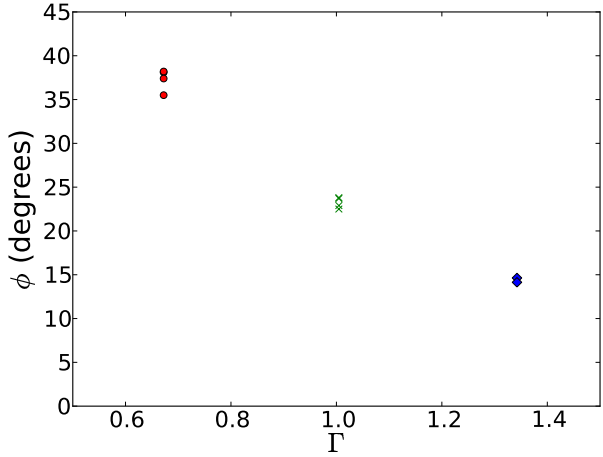}
\caption[]{The mode pitch angles for the fiducial set of simulations, R (red circles), F (green crosses) and K (blue diamonds) as a function of shear rate.}
\label{mpafid}
\end{center}
\end{figure} 

\section{Results of Fiducial Simulations}

First, we show the results of three fiducial simulations, R, F and K in Table. 1, which represent rising, flat and decreasing rotation curves respectively. In the next section, we will show results of the other simulations in Table 1 to examine the robustness of the relation between pitch angle and the shear rate shown in this section.

\subsection{Pitch angle of the mode patterns}

\begin{table} 
\centering
\begin{tabular}{c c c c}
  \hline\hline
  Simulation & $\Omega _p^m$ ($\rm{km s^{-1} kpc^{-1}}$) & \hspace{0.2cm} $m$ \hspace{0.2cm} &  \hspace{0.3cm} $\phi$ ($^{\circ}$) \hspace{0.3cm} \\ 
   \hline
   F & $ 30 $ & 3 & 23.7  \\
      & $ 42 $ &  3 & 22.9 \\ 
      & $ 35 $ &  4  & 23.8 \\ 
      & $ 40 $ &  4 & 22.5 \\     
   Fa & $35 $ & 4 & 21.2 \\
         & $45 $ & 4 & 22.4 \\
         & $28$ & 4 & 23.2 \\
   Fb & $37$ & 4 & 21.5 \\
         & $30$ & 4 & 21.6 \\
         & $45$ & 4 & 22.2 \\
   Fc & $42$ & 4 & 24.1 \\ 
        & $35$  & 4 & 22.3 \\   
        & $35$  & 5 & 24.0 \\
   F2 & $40 $ & 7 & 24.6 \\ 
   F3 & $25 $ & 3 & 26.0 \\
         & $35 $ & 3 & 25.3 \\
   K & $24 $ & 2 & 14.6 \\
       & $12 $ & 2 & 14.1 \\
   R    & $18 $ & 8 & 38.2 \\
          & $15 $ & 8 & 38.1 \\
          & $20 $ & 7 & 37.4 \\ 
          & $17 $ & 7 & 35.5 \\                             
   R2 & $30 $ & 5 & 27.8 \\
         & $25 $ & 5 & 28.7 \\
   R3 & $37 $ & 4 & 32.5 \\
         & $30 $ & 4 & 30.8 \\
         & $45 $ & 4 & 35.1 \\                
   R4 & $35 $ & 3 & 36.2 \\
         & $25 $ & 3 & 32.6 \\
   \hline
\end{tabular} 
\caption[with $65\%$ cut]{Table of mode pitch angles calculated for each simulation from the modal analysis of section 3.1. Column (1) simulation name (2) pattern speed (3) wave harmonic (4) mode pattern pitch angle.}
\label{mpavals}
\end{table}

\begin{figure*}
\begin{center}

  \subfloat{\includegraphics[scale=0.2]{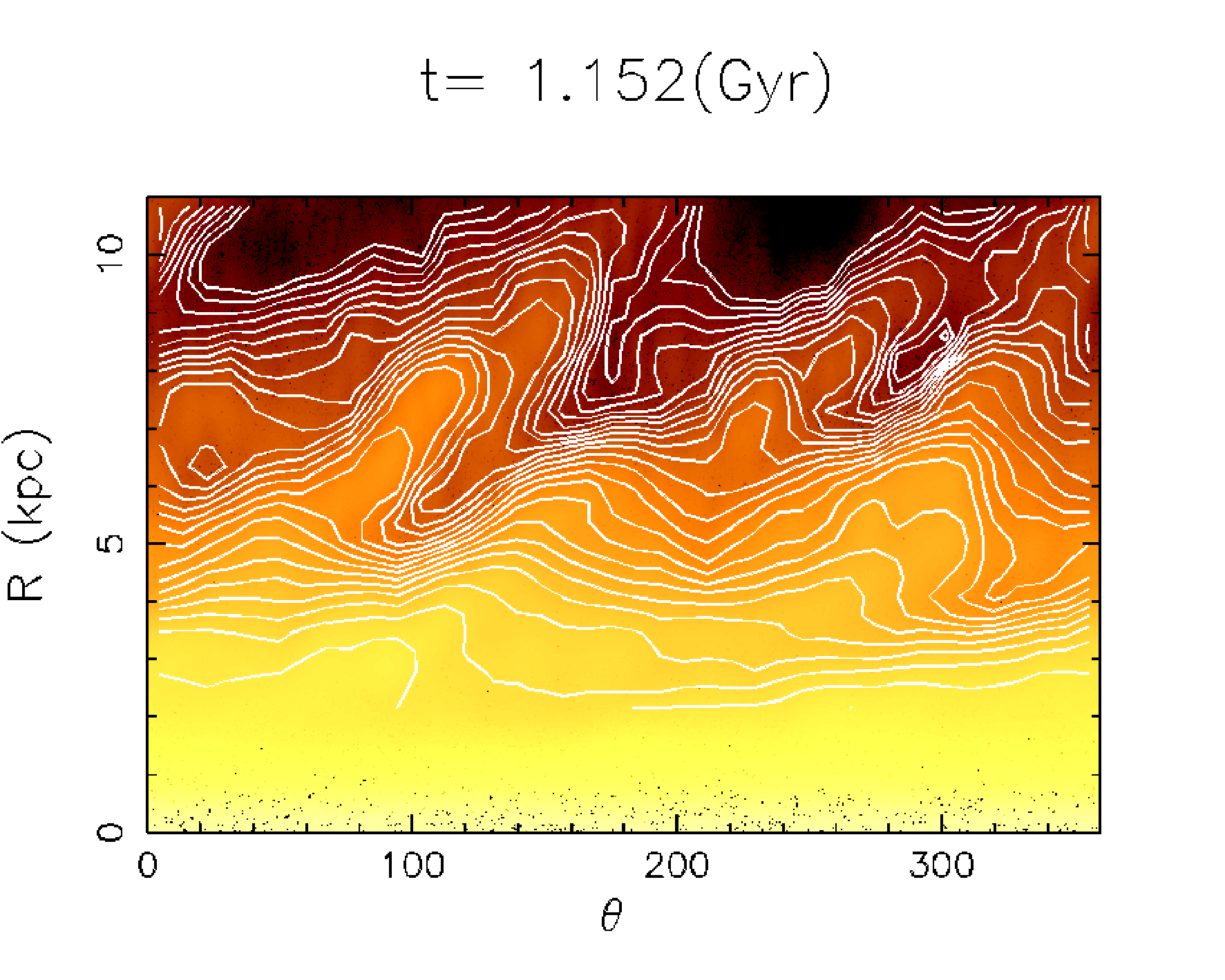}} 
  \subfloat{\includegraphics[scale=0.2] {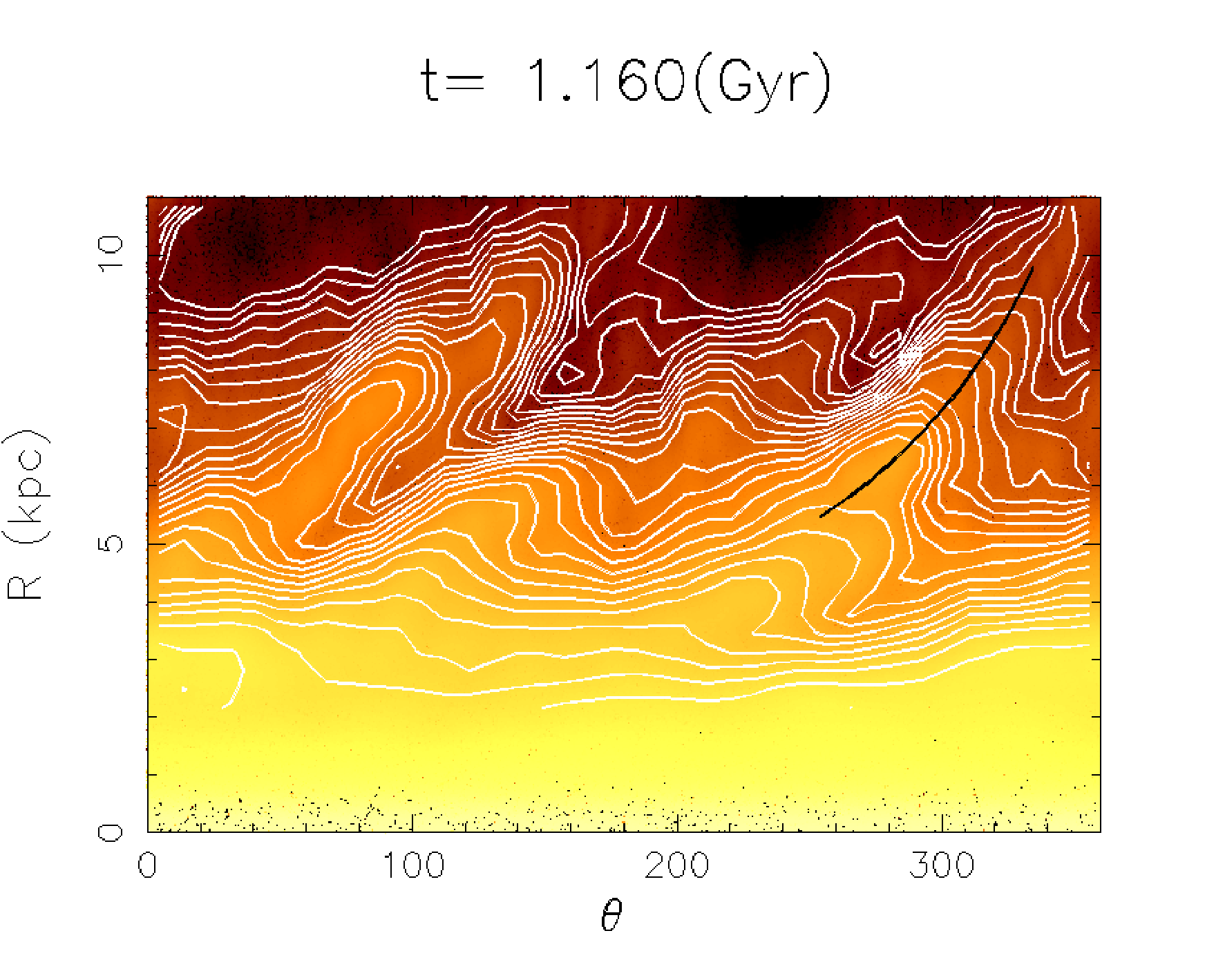}} 
  \subfloat{\includegraphics[scale=0.2] {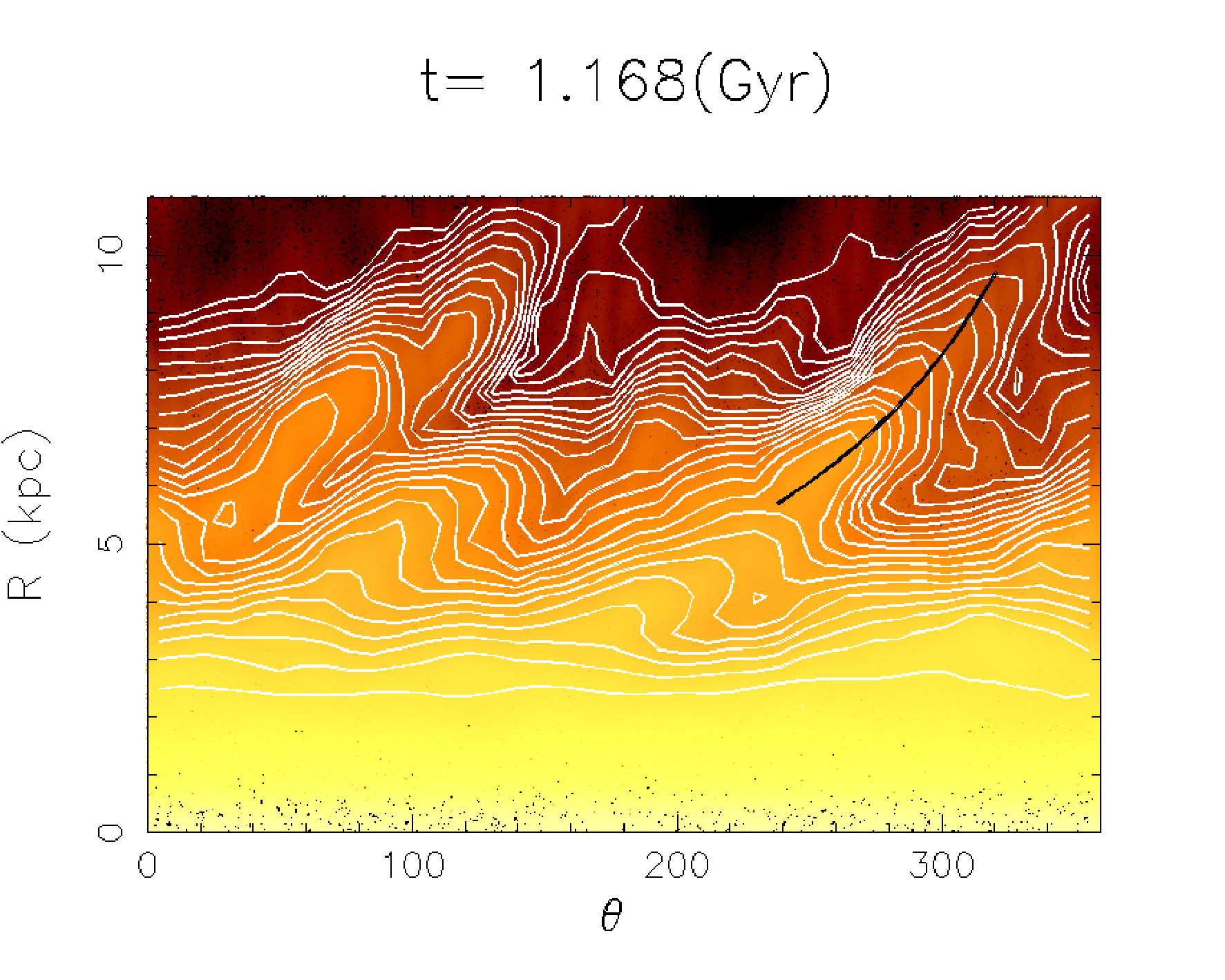}} \\
 \subfloat{\includegraphics[scale=0.2]{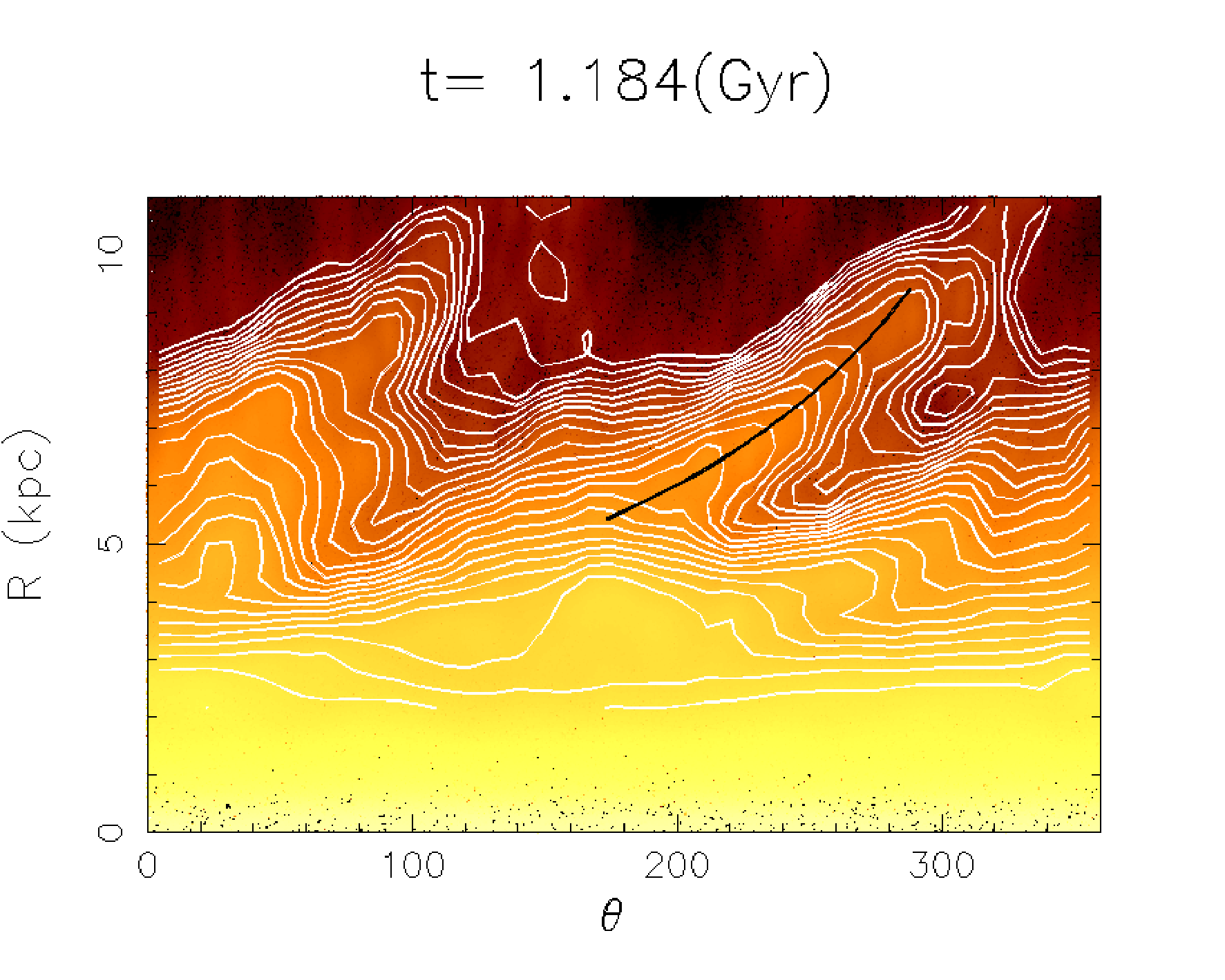}}
   \subfloat{\includegraphics[scale=0.2] {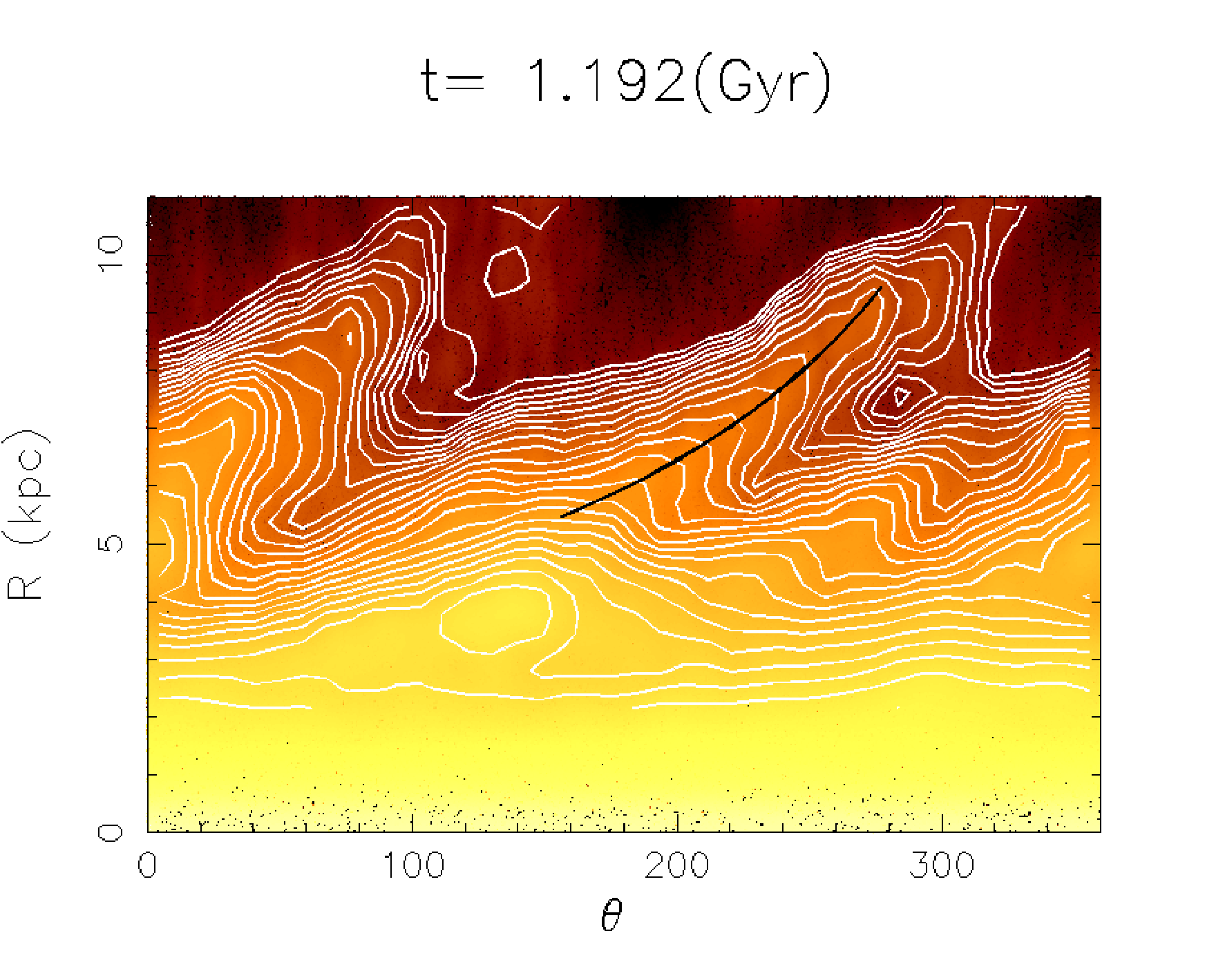}} 
  \subfloat{\includegraphics[scale=0.2] {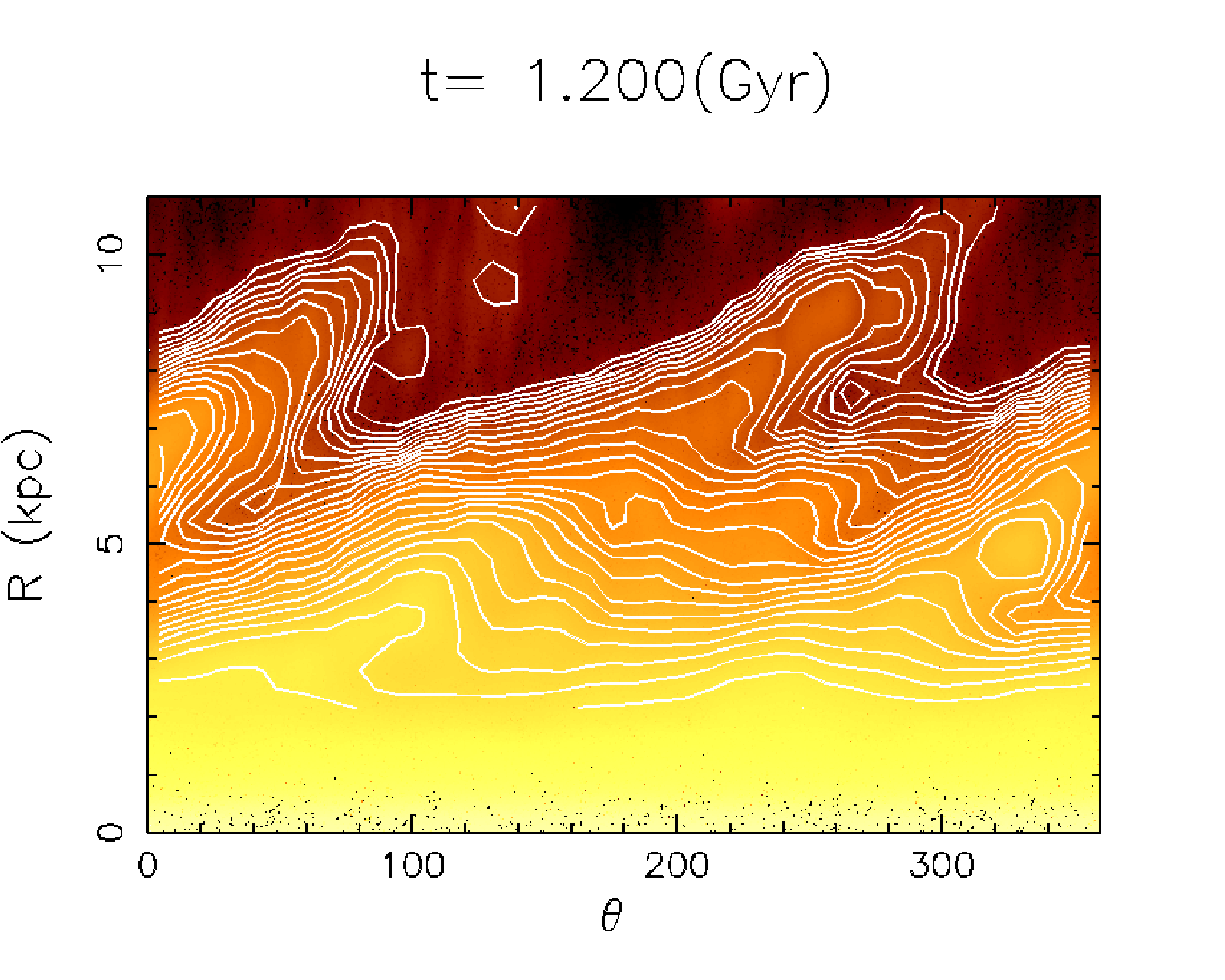}} \\

\caption[]{Snapshots of the disc density in polar coordinates. Density contours are overlaid in white. The traced spiral arm position is highlighted with a black line. The double peak structure at $R \sim 5.5$ and $\sim 9$ kpc at snapshots $t = 1.152$ and $t = 1.2$ Gyr prevents an unambiguous fitting to a single peak, and this defines the time range in which the spiral arm can be traced.}
\label{armtr}
\end{center}
\end{figure*} 

The amplitude for several wave modes is shown for each of the fiducial simulations R, F and K as a function of time in the top row of Fig. \ref{pspec}. $A_m$ is normalised to the axisymmetric amplitude, $A_0$, and averaged over the radial range $4-10$ kpc, which defines the region of spiral structure. The strong mode patterns are isolated by the vertical dashed lines in the top row of Fig. \ref{pspec}, which define the time window for the Fourier transform. The time window used is $\Delta T = 256$ Myr. Because the top row of Fig. \ref{pspec} shows that wave mode patterns appear to grow and fade on this time scale, this time window length enables the isolation of individual wave mode patterns. Although this results in limited frequency resolution, the positional information will be more reliable than that calculated from longer time windows, which may convolve multiple patterns in the Fourier analysis. However, we have confirmed that the use of longer time windows has a negligible effect on the pitch angle values. 

For each of our fiducial simulations, the power spectrum of the dominant mode highlighted in the top row of Fig. \ref{pspec} is calculated from the square of the amplitude given in equation (\ref{Ampli}), and shown as a function of radius and the pattern speed, $\Omega ^m _p = \omega /m$, in the second row of Fig. \ref{pspec}. A wave mode pattern is eligible to be analysed if its maximum power, $P^m_{max}$, is greater than $50\%$ of the maximum power of the strongest pattern, $P^{m}_{max,strongest}$ (i.e. $P^m_{max} > 0.5 P^{m}_{max,strongest}$): all other patterns are considered subsidiary. There are typically several patterns in each simulation that fulfil this criterion.

To demonstrate the fitting process, we focus on the most dominant patterns in each of the simulations R, F and K. The density maps of these dominant wave mode patterns in real space polar coordinates are shown in the third row of Fig. \ref{pspec}. This is calculated from a sinusoidal wave of the form: $\rho = A^m(R) [\cos(m(\theta - \theta _p(R))) + \sin(m(\theta - \theta _p(R)))]$. The amplitudes and phases of each radial bin are calculated from the power spectrum in the second row of Fig. \ref{pspec} using equations (\ref{Ampli}) and (\ref{Sphase}) respectively. Grey scale images highlight positive (black) and negative (white) normalised density, and contours emphasise the high density regions. The plots show coherent spiral structure with well defined pitch angles where the density contrast is high.

The bottom row of Fig. \ref{pspec} shows the logarithmic chi-squared fitting of the most dominant patterns in each simulation. The right side of each panel shows the normalised pattern amplitude as a function of radius, which reflects the relative strength of a pattern at a given radius. The logarithmic fitting is weighted by the amplitude shown in the right panel, and is represented by the red line (left panel). The fits are satisfactory for the radial ranges where the patterns are strong, and produce reliable pitch angles. The fitting of all other selected patterns for these simulations are very similar to those shown in the bottom row of Fig. \ref{pspec}. The derived pitch angles are given in Table 2.

Fig. \ref{mpafid} shows the pitch angle dependence with shear rate (equation \ref{sheqn}). All the pitch angle values clearly show a dependence on shear rate. Simulations with higher shear rate show smaller pitch angles. This is in accordance with the qualitative trend expected of the pitch angle-shear relation from theoretical studies \citep[e.g.][]{LS64,JT66}. It is interesting to note that modes of different $m$ and different pattern speeds in the same simulation (e.g. $m=3$ and $4$ in simulation F) show similar pitch angles.

\subsection{Direct pitch angles of overall spiral arm features}

As described in Section 3.2, we trace the evolution of the overall spiral arm feature directly by use of the particle density weighting method. Fig. \ref{armtr} demonstrates an example of the application of the arm tracing criteria to one of the spiral arms in simulation K. Because it is possible to reliably trace spiral arms which show only single peak structure for the radial range considered for fitting, we reject those snapshots that show the spiral arm with indistinct or double peak structure, which typically occurs during spiral arm formation ($t = 1.152$ Gyr in Fig. \ref{armtr}) and after the arm shows bifurcation or breaking ($t = 1.2$ Gyr in Fig. \ref{armtr}). 

\begin{figure}
\begin{center}
\includegraphics[scale=0.45]{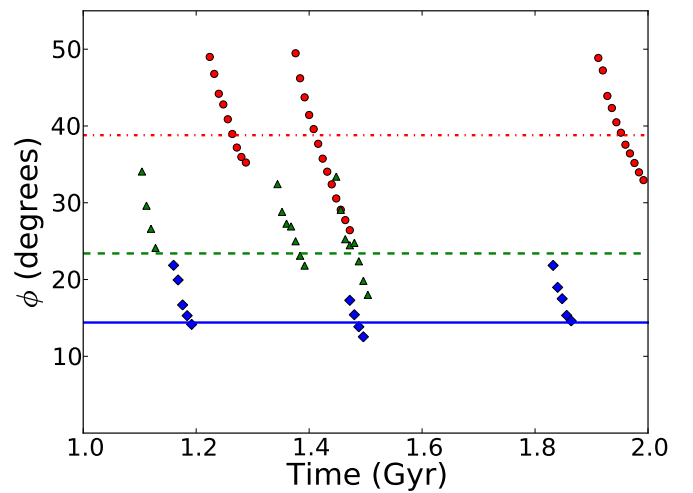}
\caption[]{Pitch angle evolution of the overall spiral arm feature for simulations R (red circles), F (green triangles) and K (blue diamonds). In all cases the pitch angle decreases with time, which indicates the winding nature of the overall density peak. The horizontal lines represent the mean mode pattern pitch angle, determined from the patterns in Fig. \ref{pspec} and shown in Table 2 for simulations R (dot-dashed red), F (dashed green) and K (solid blue). Note that the range of directly measured spiral arm pitch angles clearly map out separate domains about the mode pattern pitch angles of their respective galaxies.}
\label{AppPA}
\end{center}
\end{figure} 

\begin{figure*}
\begin{center}
  \subfloat{\includegraphics[scale=0.28]{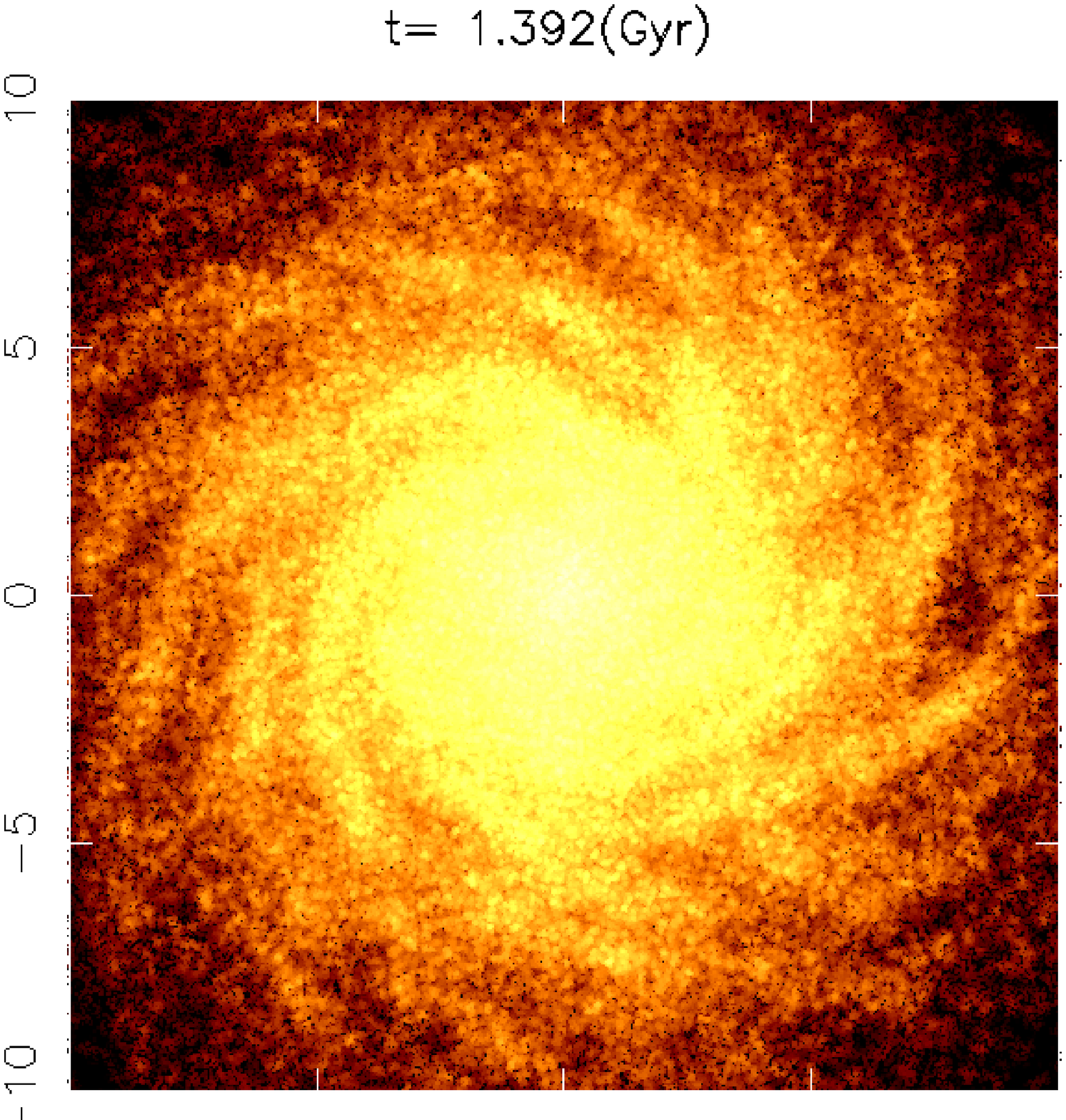}} 
  \subfloat{\includegraphics[scale=0.28] {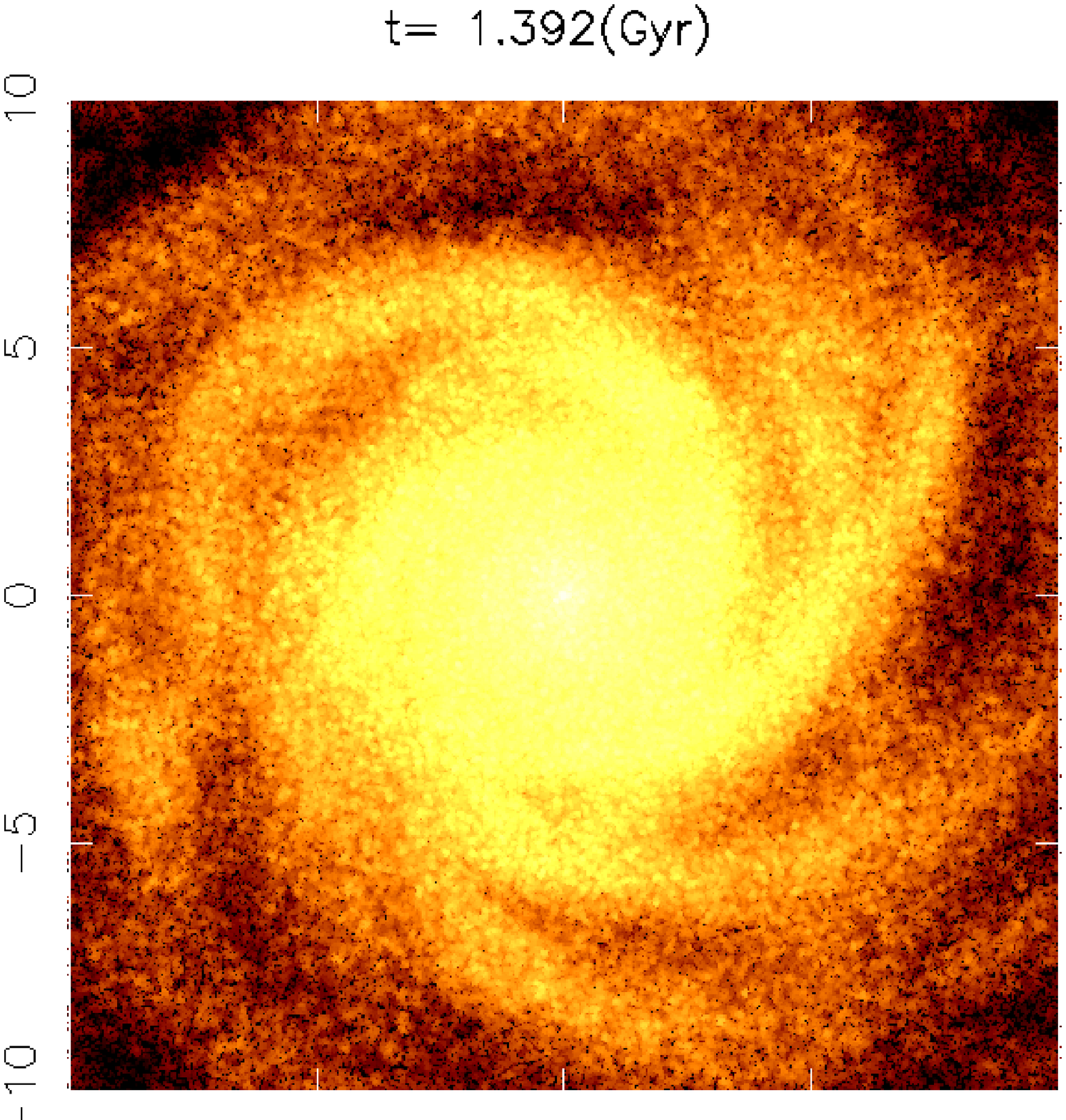}} 
  \subfloat{\includegraphics[scale=0.28] {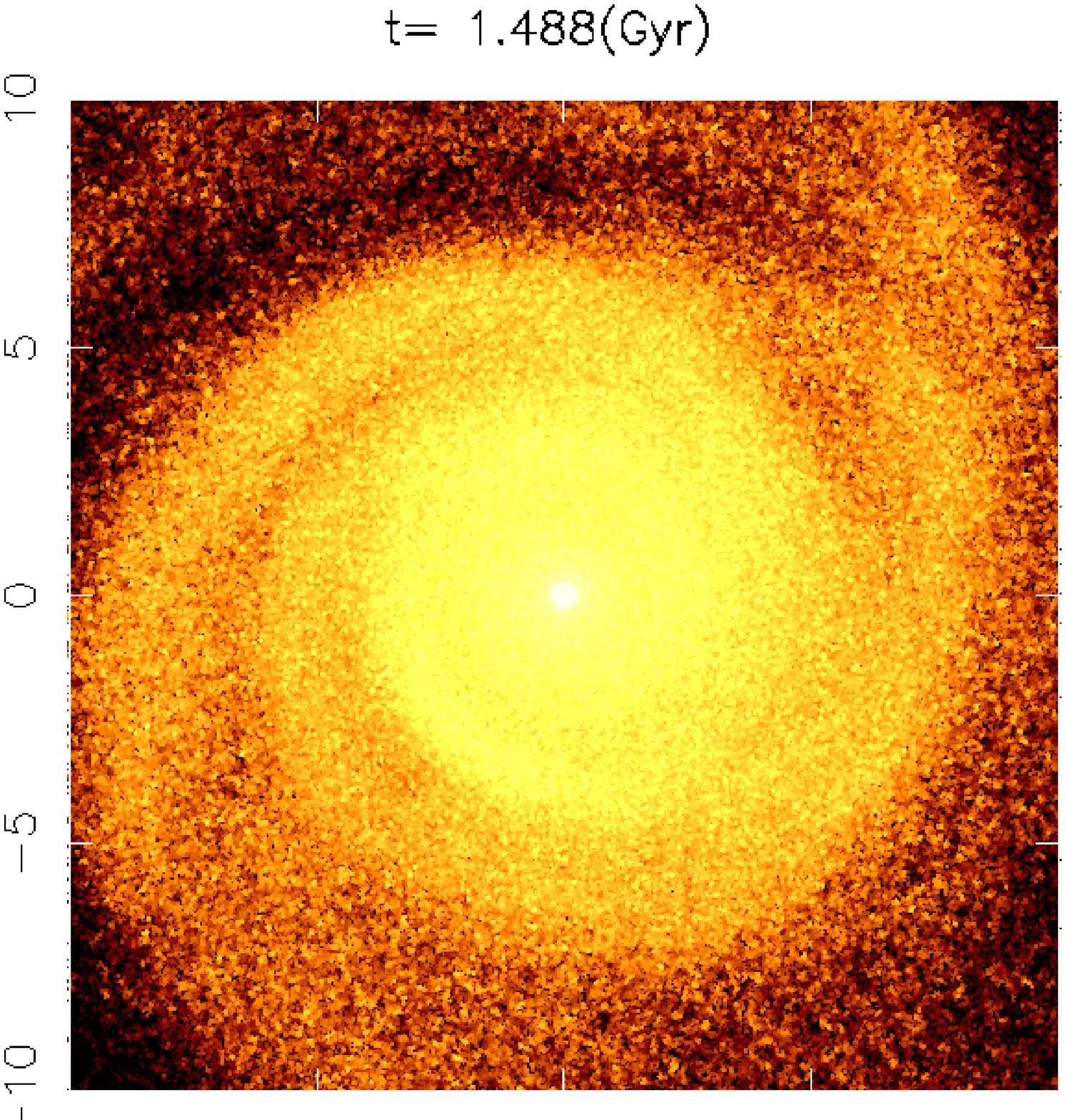}} \\

\caption[]{Face on view of each simulation (from left to right: simulations R, F and K) when the directly measured spiral arm pitch angle coincides with the calculated mode pattern pitch angle. The spirals become increasingly tight going from left to right.}
\label{faceon}
\end{center}
\end{figure*} 

The results for several spiral arms in each fiducial simulation are shown in Fig. \ref{AppPA}. It is clear that every spiral arm pitch angle decreases with time, which is consistent with winding, co-rotating spiral arms which have been reported in \citet{WBS11}, \citet{GKC12, GKC11} and \citet{BSW12}. Note that this winding is also seen in the previous formalism with mode analysis, but only through a superposition of the different mode patterns: the individual mode patterns of course are defined as being formed of fixed pattern speed, $\Omega ^m _p$, at all radii of interest. The mean of the mode pattern pitch angles calculated in the previous section is highlighted by the horizontal lines in Fig. \ref{AppPA}. The direct pitch angle values follow the same trend with shear rate as the mode pattern pitch angles presented in Section 4.1, but simulations of different shear rate can overlap in direct pitch angle owing to the spread in pitch angle values produced by the winding mechanism of the spiral arm features. A snapshot of a time when direct and mode pattern pitch angles are approximately the same is shown in Fig. \ref{faceon} for simulation R, F and K. This shows the pitch angle - shear trend clearly\footnote{Spiral arms of small pitch angle are noticed in a disc model with a massive bulge in \citet{MCB12}, who use an adaptive mesh refinement code, RAMSES \citep{T02}.}.

\begin{figure}
\begin{center}
\includegraphics[scale=0.45]{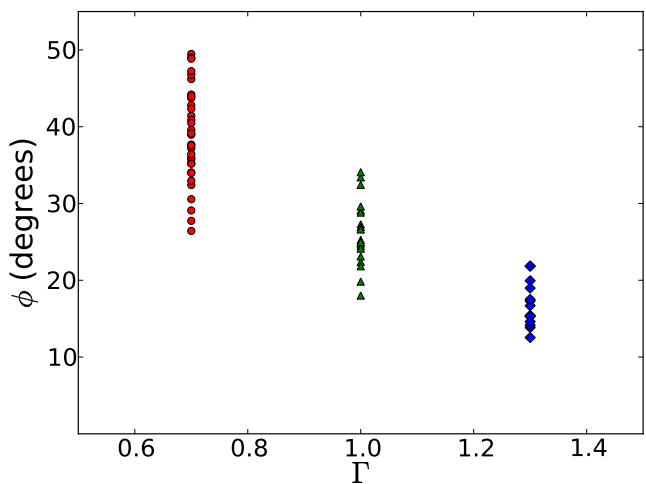}
\caption[]{All directly calculated spiral arm feature pitch angles plotted as a function of galactic shear for simulations R (red circles), F (green triangles) and K (blue diamonds).}
\label{pavsshear}
\end{center}
\end{figure} 

The winding nature of the spiral arms means that each spiral arm can exhibit several pitch angles over the spiral arm lifetime. Fig. \ref{pavsshear} shows these pitch angles plotted against galactic shear, which clearly shows that the pitch angle decreases for increasing shear rate. The range of pitch angles becomes smaller with increasing shear rate as well. This trend and scatter shown in Fig. \ref{pavsshear} are both consistent with the pitch angle-shear rate correlation and scatter seen in real observations \citep[e.g. Fig. 3 of][]{SBB06}. This may indicate that observers are seeing spiral arms at varying stages of their evolution, and therefore detect a range of pitch angles at a given rate of shear.  To test the validity of these results, we explore the effect of other parameters on pitch angle in the next section.

\section{Parameter Survey}

Up to this point, we have presented results only from the fiducial simulations R, F and K, which clearly show the relationship between pitch angle and shear rate owing to their very different rates of shear. We now explore the effects on the pitch angle of the other parameters that vary between them.

\subsection{Resolution and Softening length}

\begin{figure*}
\begin{center}

  \subfloat{\includegraphics[scale=0.3]{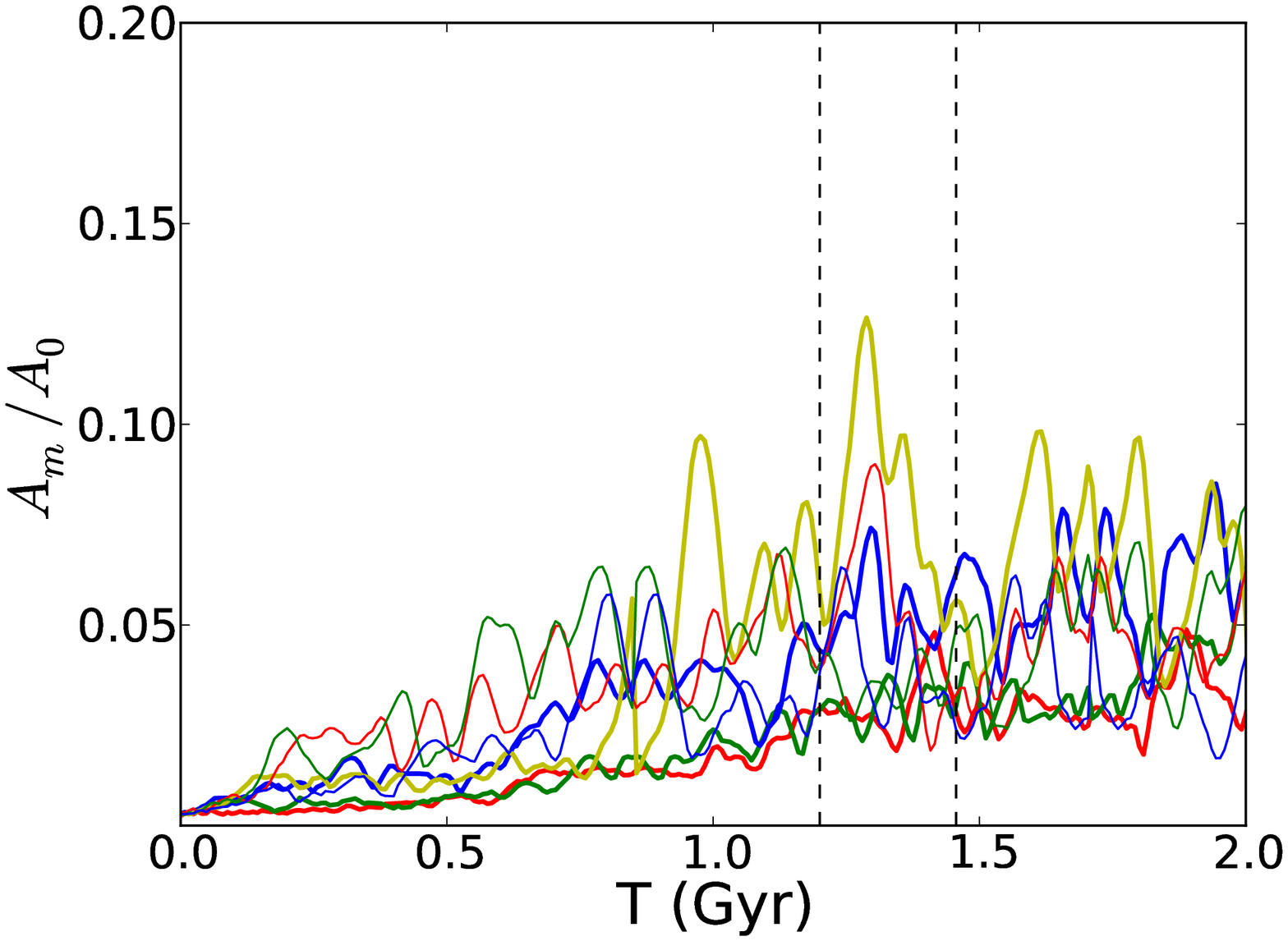}} 
  \subfloat{\includegraphics[scale=0.3] {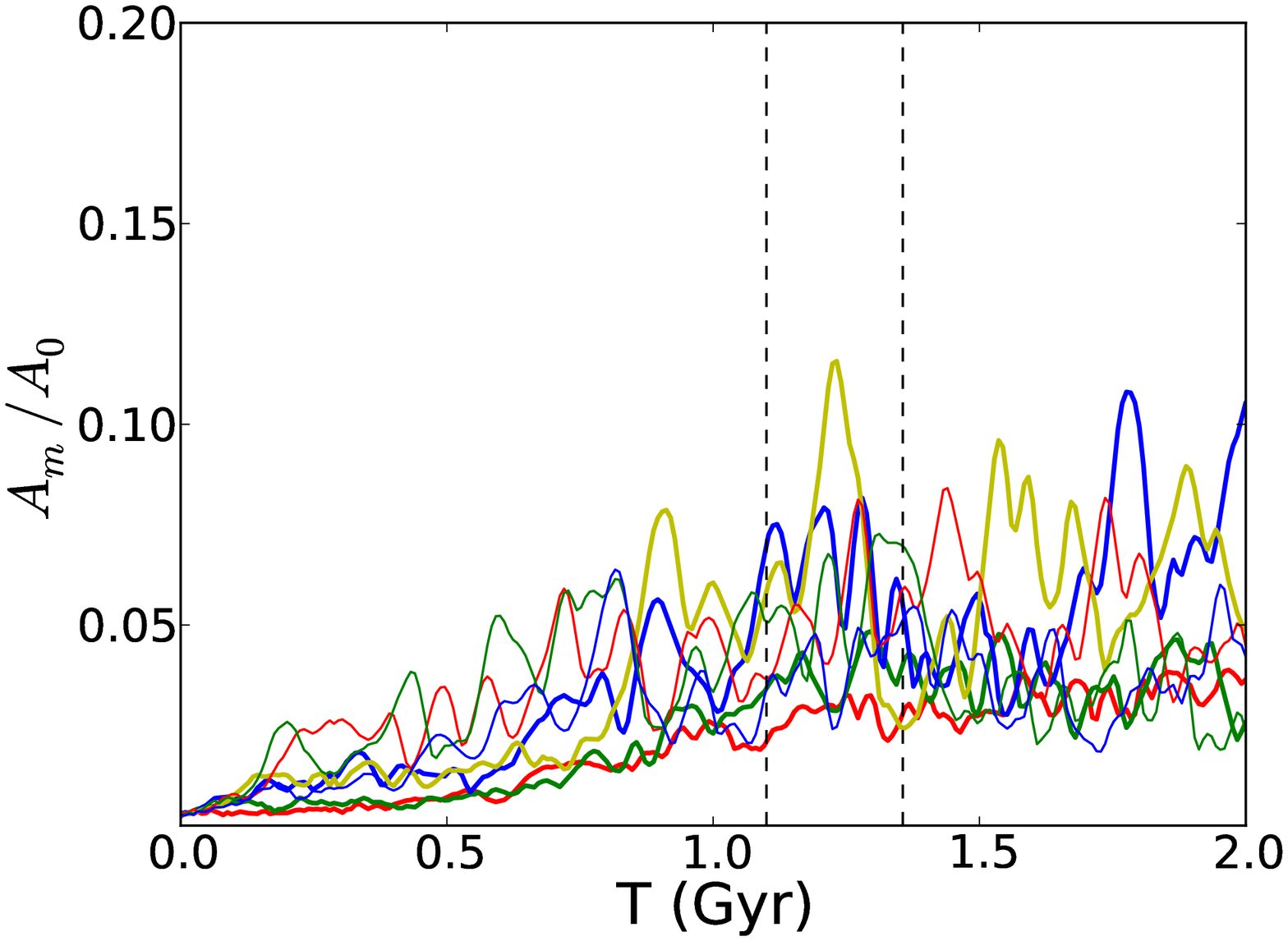}} 
  \subfloat{\includegraphics[scale=0.3] {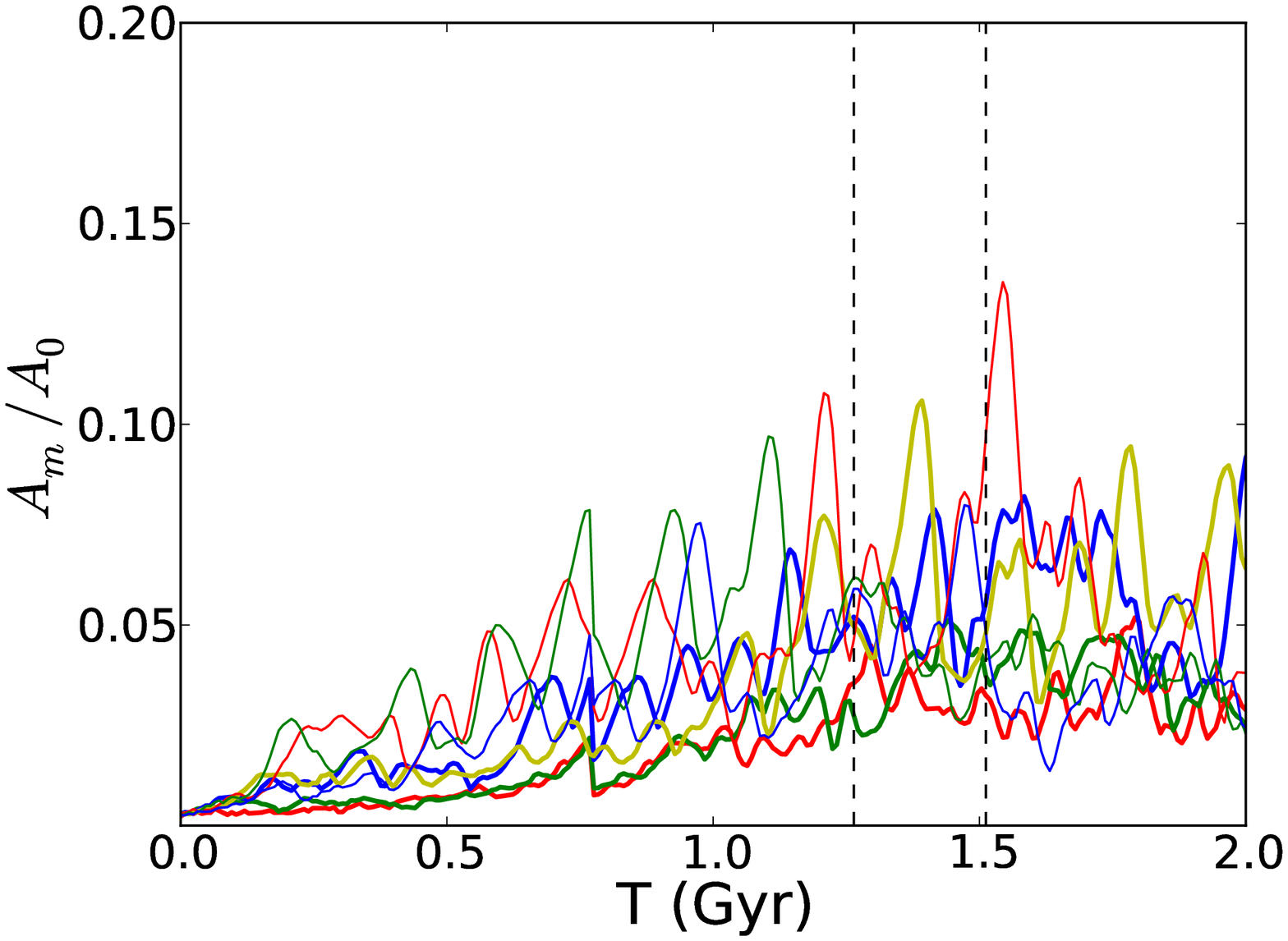}} \\
 \subfloat{\includegraphics[scale=0.3]{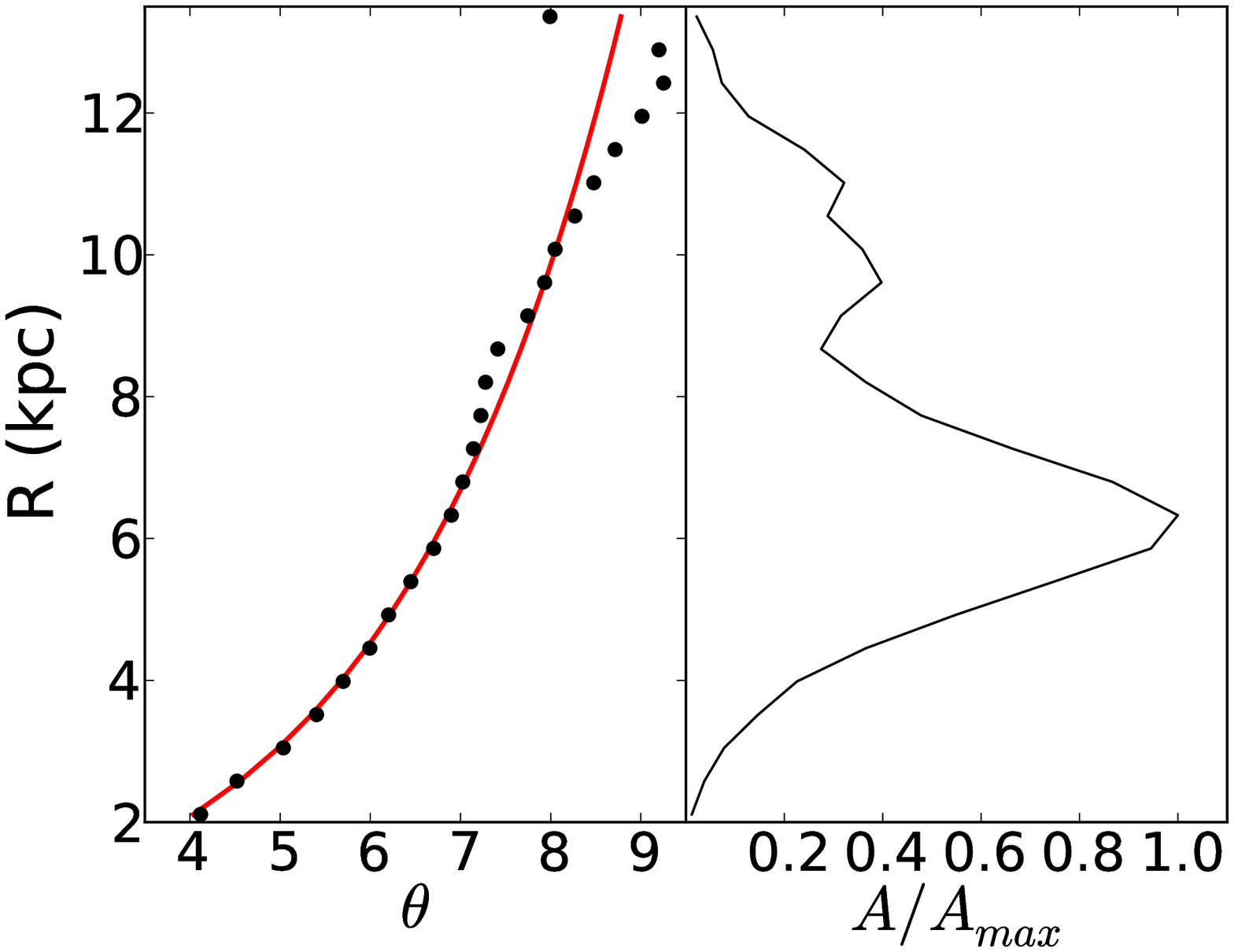}}
   \subfloat{\includegraphics[scale=0.3] {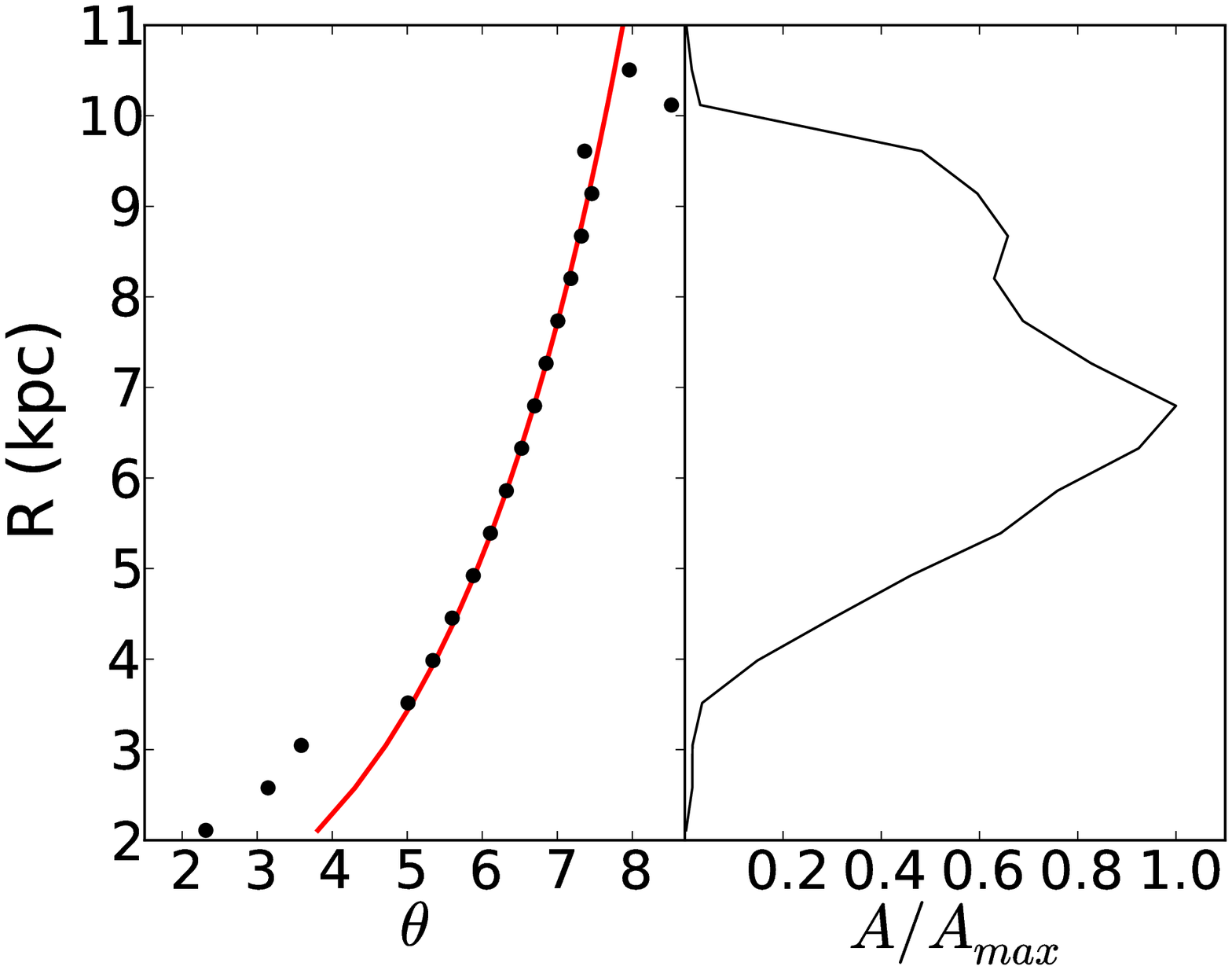}} 
  \subfloat{\includegraphics[scale=0.3] {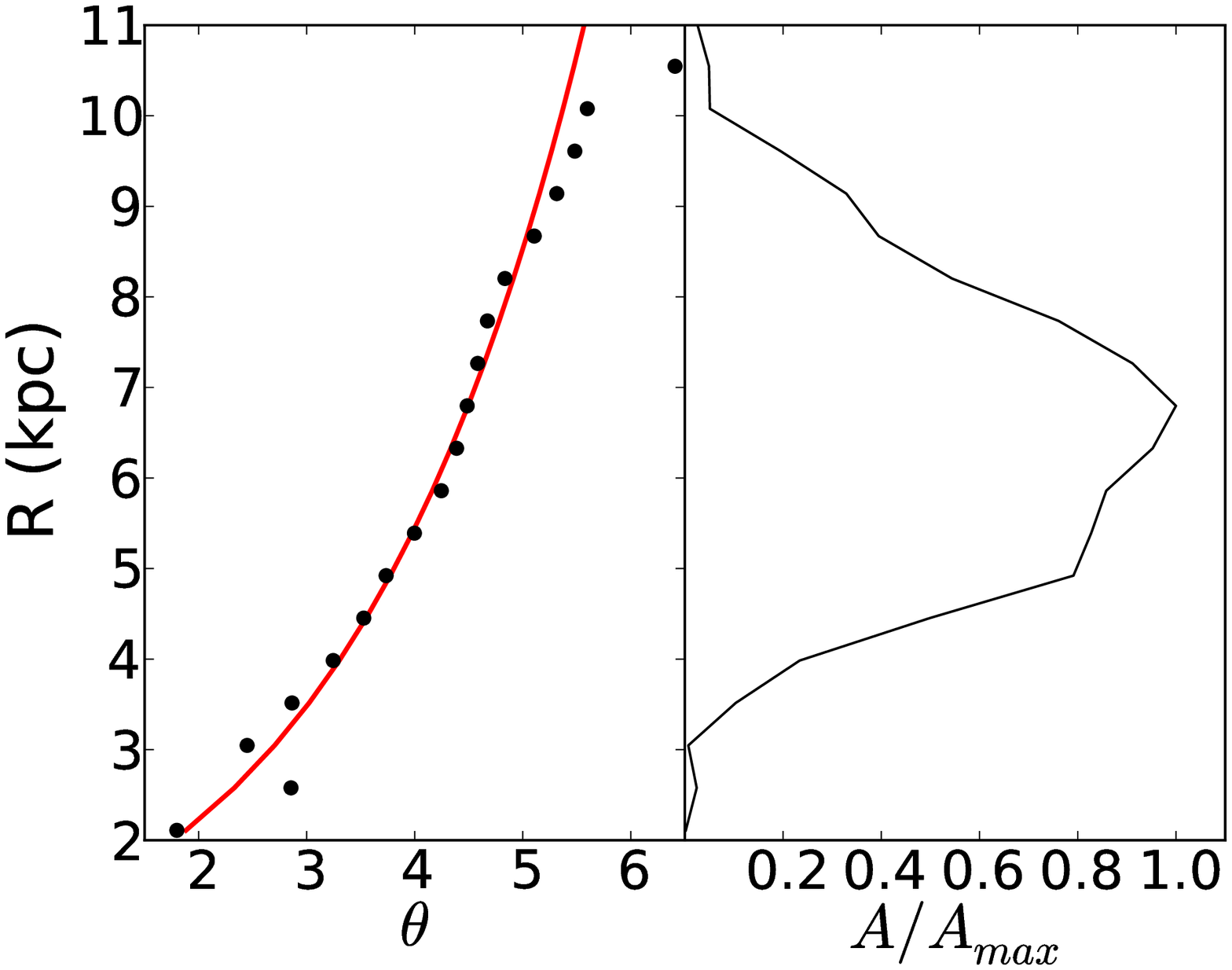}} \\

\caption[]{Top row: Amplitudes of the $m=1-7$ wave mode numbers (colours as in top row of Fig. \ref{pspec}). Bottom row: Phase positions of the strong mode patterns identified in top row. From left to right: simulations Fa ($m=4$), Fb ($m=4$) and Fc ($m=4$) respectively. }
\label{rescomp}
\end{center}
\end{figure*}

We investigate the numerical robustness of the simulations by examining the effect of the number of particles and the choice of softening length. We start with simulations Fa, Fb and Fc, which use N=$5 \times 10^6$ particles with different softening lengths (see Table 1) together with the fiducial F. They are identical in every other parameter to the fiducial F simulation. The top row of Fig. \ref{rescomp} shows their wave mode amplitudes and dominant mode pattern phase positions. There are some differences between the higher resolution simulations, Fa, Fb and Fc. For example, the $m=5$ mode shows significant amplitude in Fc.   

Because the softening length relates to the particle mass as $\epsilon \propto m_p^{1/3}$, a direct comparison to explore the effect of resolution is between simulation F and Fb. The spiral structure grows slightly more slowly in simulation Fb (as well as the other higher resolution simulations) than in simulation F, but modes of $m=3$ and $4$ remain strong in all of these simulations. The difference in level of spiral structure growth for the different particle number is as expected \citep{Fu11}. 

The chi-squared fitting of the most dominant patterns in each simulation is shown in the bottom row of Fig. \ref{rescomp}. The mode pattern pitch angles for all three higher resolution simulations are given in Table 2, and are all very similar to the fiducial F mode pattern pitch angles.

\begin{figure}
\begin{center}
\includegraphics[scale=0.45]{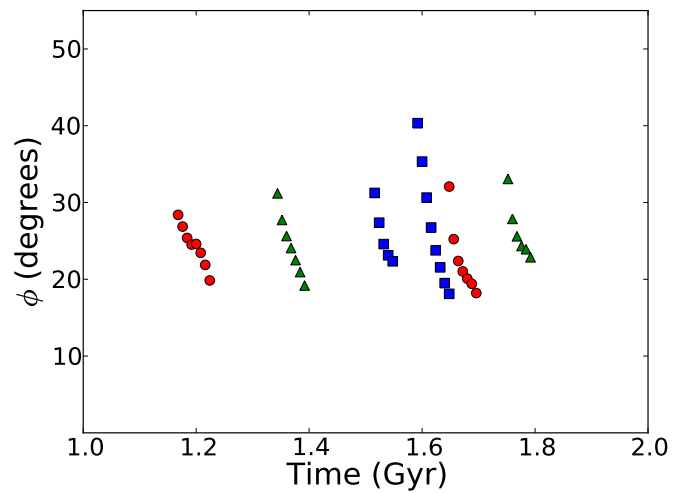}
\caption[]{As for Fig. \ref{AppPA} but for simulations Fa (blue squares), Fb (red circles) and Fc (green triangles).}
\label{compres}
\end{center}
\end{figure}

Fig. \ref{compres} shows the pitch angles of several spiral arms that we analysed using the direct trace of the spiral arm features. Again, the arms are winding with time, and the range of pitch angles are consistent with simulation F in Fig. \ref{AppPA}. In Fig. \ref{compres}, at around $t=1.6$ Gyr, simulation Fa shows a spiral arm that forms with an initial pitch angle of $\phi = 41$ degrees, and is quickly wound. Although this initial pitch angle is high compared to that of the other arms, the later pitch angle measurements for this spiral arm overlap the range of pitch angles of all the other arms in simulations F, Fa, Fb and Fc. 

The general agreement between the mode pattern pitch angles and the range of direct pitch angles over the simulations F, Fa, Fb and Fc indicates that the fiducial resolution of $N = 1$ million particles is sufficient to capture robust pitch angles. Moreover, the variation of the softening length in the assumed range does not appear to be a significant factor either, owing to the very similar mode pattern pitch angles given in Table 2 and directly measured pitch angles shown in Fig. \ref{compres}.

\subsection{Disc-Halo mass ratio}

\begin{figure*}
\begin{center}

  \subfloat{\includegraphics[scale=0.3]{Fampav}} 
  \subfloat{\includegraphics[scale=0.3] {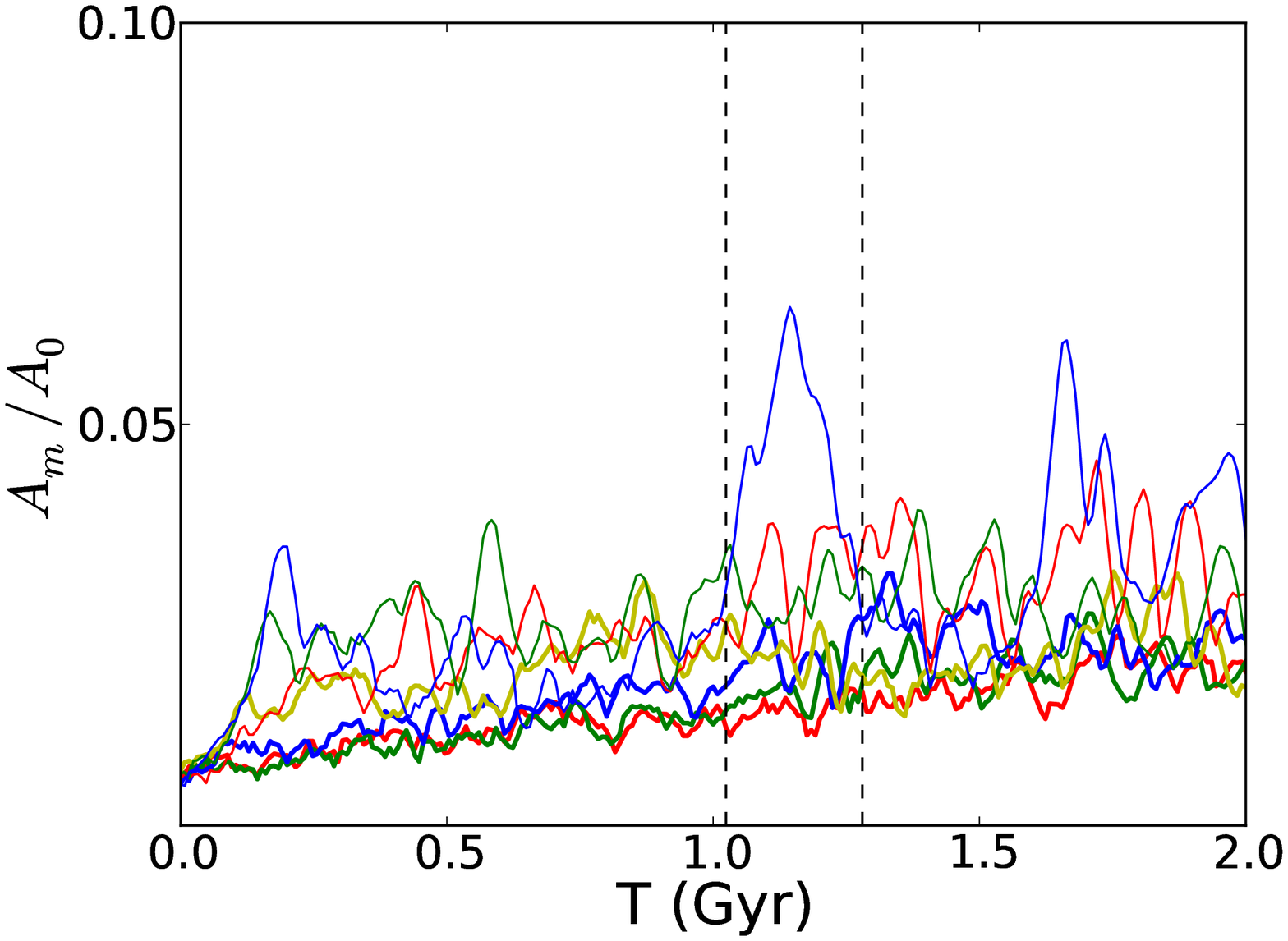}} 
  \subfloat{\includegraphics[scale=0.3] {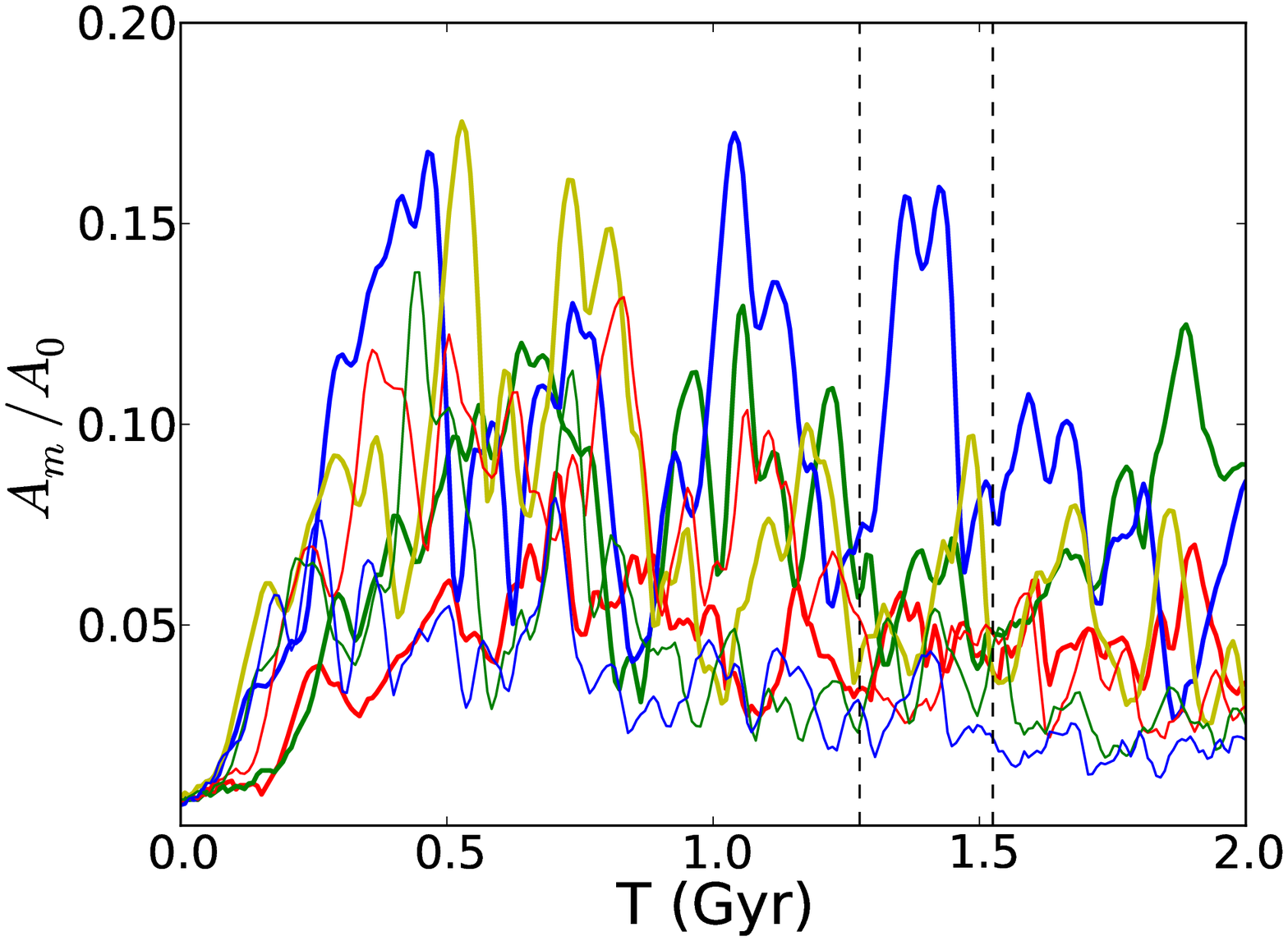}} \\
 \subfloat{\includegraphics[scale=0.3]{Ffit3mp1}}
   \subfloat{\includegraphics[scale=0.3] {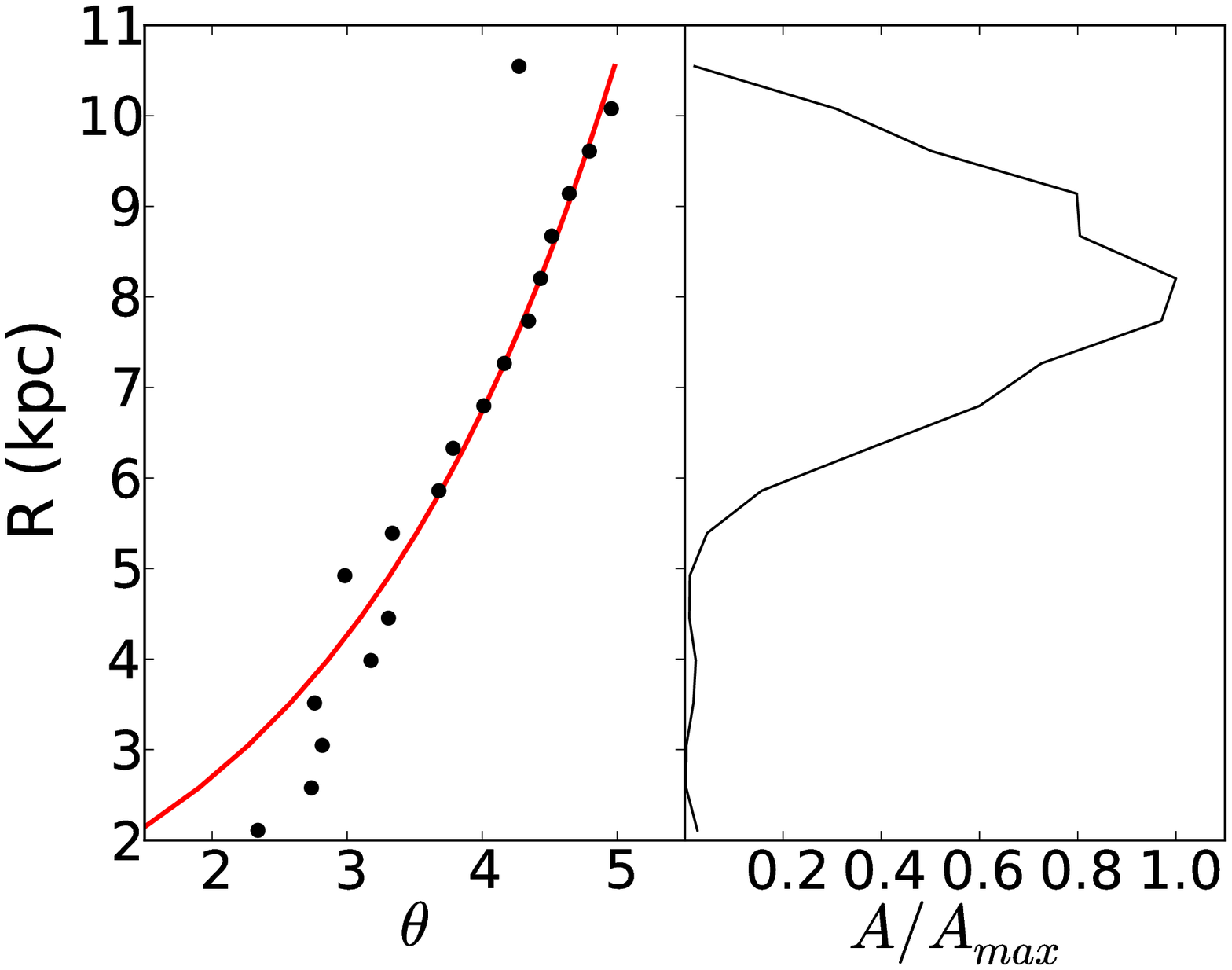}} 
  \subfloat{\includegraphics[scale=0.3] {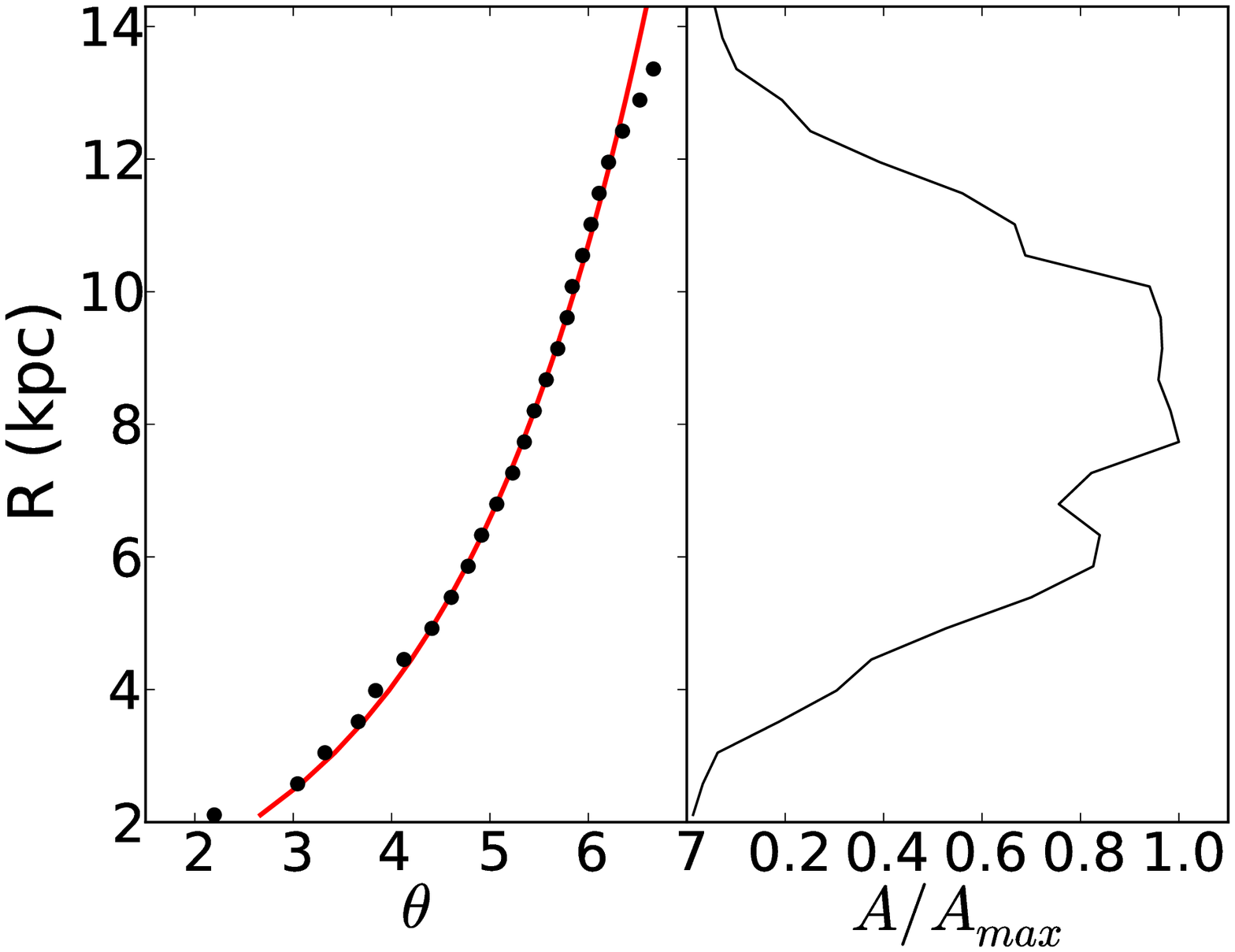}} \\

\caption[]{As in Fig. \ref{rescomp}, but for simulations F ($m=3$), F2 ($m=7$) and F3 ($m=3$).}
\label{ratiocomp}
\end{center}
\end{figure*} 

\begin{figure}
\begin{center}
\includegraphics[scale=0.45]{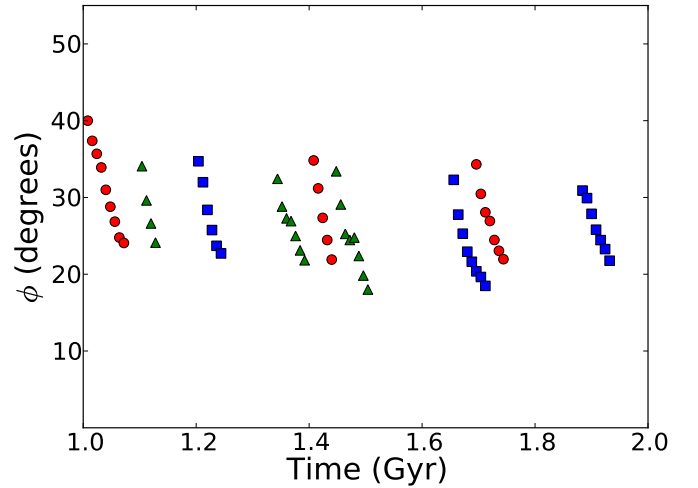}
\caption[]{As for Fig. \ref{AppPA} but for simulations F (green triangles), F2 (blue squares) and F3 (red circles).}
\label{compflat}
\end{center}
\end{figure}

Another variable in our simulations is the disc mass to halo mass ratio. To see whether or not this parameter affects the pitch angle, we perform the same analysis on simulations F2 and F3, which display shear rates within $\sim 2 \%$ of the fiducial simulation F, with lower and higher disc-halo mass ratios respectively (see Table 1). This ratio, $\zeta$, is calculated as the ratio of the disc mass to the external mass within two radial scale lengths \citep[as performed in][]{DO12}. The amplitudes and density mode pattern phase positions are shown in Fig. \ref{ratiocomp}. The mode pattern pitch angles calculated from the fitting in the bottom rows in Fig. \ref{ratiocomp} is presented in Table 2. The pitch angle values of F2 and F3 are similar to that of F. The directly measured pitch angles from the spiral arm feature shown in Fig. \ref{compflat} also show little difference between the simulations, with perhaps the exception of the F3 spiral arm beginning $t=1$ Gyr at $\phi \sim 40 ^{\circ}$. Overall, these results indicate that the disc to halo mass ratio does not affect the pitch angle of the spiral features, but instead the number of spiral arms, $m$. For example, in Fig. \ref{ratiocomp} the higher disc-mass ratio simulation, F3, displays more power in lower wave mode numbers ($m = 2, 3$) whereas the lowest disc-halo mass ratio simulation, F2,  shows the $m=7$ mode to be most prominent. This is consistent with previous studies \citep{JT66,T81,ELN82,CF85,DO12}.

\begin{figure}
\begin{center}
\includegraphics[scale=0.45]{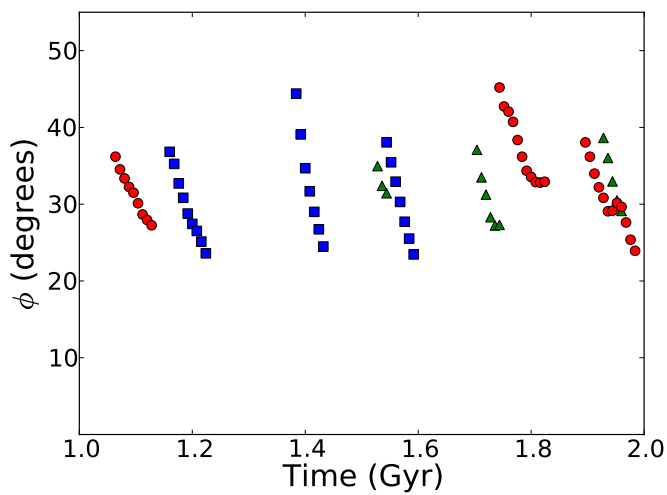}
\caption[]{As for Fig. \ref{AppPA} but for simulations R2 (blue squares), R3 (green triangles) and R4 (red circles).}
\label{compris}
\end{center}
\end{figure}

We also performed simulations of intermediate shear rate values between simulations R and F with a slight alteration of disc-halo mass ratio. These simulations, labelled R2, R3 and R4 (in order from higher to lower shear), have no bulge. Fig. \ref{compris} shows the direct pitch angle of several spiral arms in these simulations. While they are similar to each other, the range of pitch angles covers a slightly higher range than that of simulation F but slightly lower than that of simulation R. This agrees with the intermediate shear values shown in Fig. \ref{shear}. Table 2 shows the measured pitch angle of the wave modes, which also indicates the intermediate mode pattern pitch angles between simulations R and F. 

\begin{figure}
\begin{center}
\includegraphics[scale=0.45]{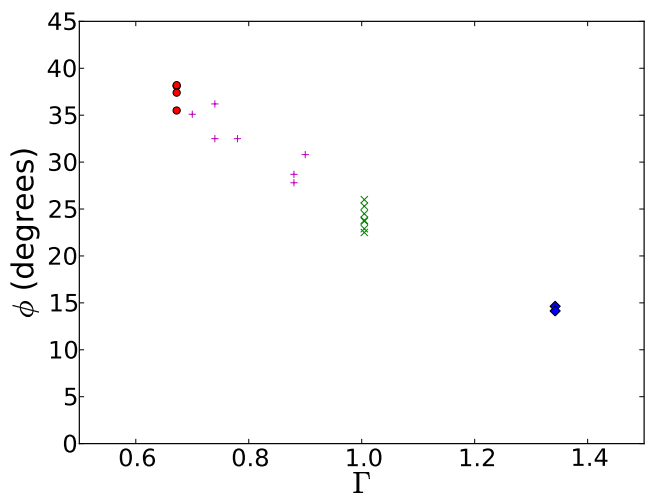}
\caption[]{The mode pitch angles as a function of shear for simulations, R (red circles), R2, R3, R4 (magenta plusses), F, F2, F5 (green crosses) and K (blue diamonds).}
\label{mainfig}
\end{center}
\end{figure}

To examine the trends together, we plot the mode pattern pitch angles of simulations F, F2, F3, K, R, R2, R3 and R4 as a function of shear rate in Fig. \ref{mainfig}. This figure shows a clear correlation between pitch angle and shear rate, which is the main finding of this paper.

The lack of effect of disc-halo mass ratio on pitch angle in combination with the difference in pitch angle between simulations F and R, which both have the same mass ratio, are convincing evidence that the shear rate is the dominant driver of pitch angle in N-body simulations of spiral galaxies.

\section{Discussion}

We have shown that in N-body simulations, the measured pitch angles (measured both through the wave mode patterns and directly tracing the spiral arm features) correlates with shear rate. The range of direct pitch angles produced is in agreement with observation. We explored other simulation parameters, and show that the pitch angle is not significantly affected by the disc-halo mass ratio, resolution or softening length. One other parameter whose effect we could not explore is the stability parameter, $Q$, owing to the fact that it cannot be directly specified and it evolves over time \citep{Fu11}.  Although we could not test this parameter directly, we note that the $Q$ parameter is reported from analytical studies \citep[e.g.][]{JT66,A84,F01} to have negligible effect on the pitch angle of swing-amplified patches. Also, the density wave theory of \citet{LS64} does not show an explicit correlation between the pitch angle and the $Q$ parameter. Therefore, we expect the major driver of the pitch angle value of spiral arms in N-body simulations to be the shear rate. However, this aspect still needs further study. 

The observed correlation between the pitch angle of the density wave mode and galactic shear rate is qualitatively consistent with the prediction of the classic theories of both density wave theory \citep{LS64} and swing amplification theory \citep{JT66,T81}. 

In the context of swing amplification theory, spiral structure grows from density perturbations as the stellar material swings from an open to a tightly wound structure, so as to exhibit a range of inclination angles. Therefore the pitch angle may correspond to the inclination angle when each density perturbation is most amplified, around a specific inclination angle, which is correlated to shear rate \citep{JT66}.

In the context of the Lin-Shu density wave theory, each wave mode can be interpreted as a standing wave mode of constant pitch angle and fixed pattern speed. \citet{LS64} demonstrate that the pitch angle of such waves is lower for higher central mass concentrations, i.e. a higher shear rate. However, there must be more than one wave mode to manifest the winding of the spiral arm, which must then be interpreted in terms of a superposition of multiple mode patterns, which changes with time \citep[e.g.][]{CQ12}. In this interpretation, the wave mode patterns in the inner disc region must have a faster pattern speed than that in the outer region, and must overlap at some intermediate radii. Therefore, the pitch angle begins larger than that measured for the wave mode, and then approaches the mode pitch angle while the density grows (constructive interfering). The waves then pass and move away from one another, which decreases the pitch angle further. This leads to a stretch in the azimuthal direction of the overall spiral arm density. 

If multiple wave modes are the driving mechanism of spiral arms, the N-body simulations suggest that there are many patterns of various multiplicity, $m$, that are short-lived (as seen from the transient nature of the mode amplitudes in the top row of Fig. \ref{pspec} for example) and recurrent. However, it is worth noting that such waves are some distance from the large scale, long timescale structures that classic spiral density wave theory was developed to produce. The formation and evolution of such wave modes should be non-linear and complicated \citep{DO12,BSW12}, which deserves further study, and is beyond the scope of this paper.

\section{Conclusions}

For the first time, to our knowledge, we have analysed the pitch angle of the spiral arm features directly and the pitch angle of the wave mode pattern in N-body simulations of disc galaxies with different galactic shear rate. The former pitch angle is derived from tracing the physical movement of the actual surface density of the spiral arms, and the latter is calculated from Fourier analysis that aims to isolate density wave mode patterns from the system that may contribute to the overall movement of the spiral arms. We presented and compared the results of both techniques, and come to the following conclusions.
\begin{enumerate}
\item{}
We find that the pitch angle measured both through the wave mode analysis and direct analysis is correlated with the rate of galactic shear: the pitch angle is smaller for higher galactic shear rate and vice versa. This is consistent qualitatively with the analytical predictions based on density wave theory \citep{LS64} and swing amplification theory in \citet{JT66}, which we demonstrate in N-body simulations for the first time.
\item{}
The direct pitch angles of the overall spiral arm density enhancement decrease with time, as the spiral arms grow from a relatively open arm morphology, then wind over time to become more tightly wound until they disrupt. This is consistent with previous simulations that reported winding and co-rotating spiral arms \citep{WBS11,GKC12,GKC11,BSW12}. 
\item{}
The range of the direct pitch angles resulting from the winding spiral arm features is correlated with their shear rate: the direct pitch angle range tends to be smaller for the system with higher galactic shear and vice versa. The range of direct pitch angles at a given shear rate is similar to the scatter seen from the observed relation between the pitch angle and the shear rate in spiral galaxies reported in \citet{SBB06}. This is consistent with the view that real galaxies exhibit transient and winding spiral arms.  
\end{enumerate}

Our N-body simulations demonstrate the relation between the pitch angle and the galactic shear rate. Although we explored several parameters, such as disc-total mass ratio and simulation resolution, this area of study is far from completion. We also used a fixed dark matter halo for simplicity, and left out the gas component. In real galaxies, there are also constant minor mergers and tidal interactions with satellite galaxies, which we have not explored. However, we suggest that this study highlights the relation between pitch angle and the galactic shear rate, and encourages further studies with more realistic and complicated models. If this relation is a dominant mechanism to determine the pitch angle of the spiral arms, because the late type spiral galaxies tend to have rising rotation curves, this relation will become key to explain the correlation between the pitch angle and the Hubble type \citep{H26,K81}.

\section{Acknowledgements}
The authors thank Jerry Sellwood for suggestions regarding the mode pattern analysis.
The authors acknowledge the support of the UK's Science \& Technology
Facilities Council (STFC Grant ST/H00260X/1). The calculations for
this paper were performed on Cray XT4 at Centre for
Computational Astrophysics, CfCA, of National Astronomical Observatory
of Japan and the DiRAC Facilities, Legion and COSMOS, jointly funded by STFC and the Large
Facilities Capital Fund of BIS. The authors acknowledge support of the
STFC-funded Miracle and COSMOS Consortium (part of the DiRAC facility) in
providing access to the UCL Legion and Cambridge COSMOS High Performance Computing
Facilities.  The authors additionally acknowledge the support of UCL's
Research Computing team with the use of the Legion facility. 
This work was carried out, in part, through the Gaia
Research for European Astronomy Training (GREAT-ITN) network. The
research leading to these results has received funding from the
European Union Seventh Framework Programme ([FP7/2007-2013] under
grant agreement number 264895.

\bibliographystyle{aa}
\bibliography{sheargam_feb18.bbl}

\end{document}